\documentclass[aps,prd,onecolumn,10pt,preprint,nolongbibliography]{revtex4-2}
\usepackage[export]{adjustbox}
\usepackage{bm}
\usepackage{float}
\usepackage{amsmath}
\usepackage{color}
\usepackage{graphicx}
\usepackage{listings}
\usepackage{float}
\usepackage{soul}
\restylefloat{table}

\begin{document}

\title{Potential Energy Landscape of a Flexible Water Model: Equation-of-State, Configurational Entropy, and Adam-Gibbs Relationship}

\author{Ali Eltareb$^{1,2,*}$}
\author{Gustavo E. Lopez$^{3,4,*}$}
\author{Nicolas Giovambattista$^{1,2,4}$}

\email{ali.eltareb@brooklyn.cuny.edu, gustavo.lopez1@lehman.cuny.edu, ngiovambattista@brooklyn.cuny.edu}

\affiliation{$^1$Department of Physics, Brooklyn College of the City University of New York, Brooklyn, New York 11210, United States \\
	$^2$Ph.D. Program in Physics, The Graduate Center of the City University of New York, New York, NY 10016, United States \\
	$^3$Department of Chemistry, Lehman College of the City University of New York, Bronx, New York 10468, United States \\
	$^4$Ph.D. Program in Chemistry, The Graduate Center of the City University of New York, New York, NY 10016, United States}

\begin{abstract}
The potential energy landscape (PEL) formalism is a tool within statistical mechanics that has been used in the past to calculate the 
equation of states (EOS) of classical {\it rigid} model liquids at low temperatures, where computer simulations may be challenging. In this work, we use classical molecular dynamics (MD) simulations and the PEL formalism to calculate the EOS of the {\it flexible} q-TIP4P/F water model. This model exhibits a liquid-liquid critical point (LLCP) in the supercooled regime, at ($P_c = 150$~MPa, $T_c = 190$~K, $\rho_c = 1.04$~g/cm$^3$) [using the reaction field technique]. The PEL-EOS of q-TIP4P/F water, and the corresponding location of the LLCP, are in very good agreement with the MD simulations. We show that the PEL of q-TIP4P/F water is Gaussian which allows us to calculate the configurational entropy of the system, $S_{conf}$. The $S_{conf}$ of q-TIP4P/F water is surprisingly similar to that reported previously for rigid water models, suggesting that intramolecular flexibility does not necessarily add roughness to the PEL. We also show that the Adam-Gibbs relation, which relates the diffusion coefficient $D$ with $S_{conf}$, holds for the flexible q-TIP4P/F water model. Overall, our results indicate that the PEL formalism can be used to study molecular systems that include molecular flexibility, the common case in standard force fields. This is not trivial since the introduction of large bending/stretching mode frequencies is problematic in classical statistical mechanics. For example, as shown previously, we find that such high-frequencies lead to an unphysical (negative) entropy for q-TIP4P/F water (yet the PEL formalism can be applied successfully).
\end{abstract}

\maketitle
\clearpage
\section{Introduction}

The potential energy landscape (PEL) formalism~\cite{goldstein1969viscous,stillinger1982hidden,stillinger1984packing} in classical statistical mechanics was introduced as a tool for understanding 
the structure, kinetics, and phase behavior in condensed matter~\cite{stillinger1982hidden,stillinger1984packing,goldstein1969viscous,stillinger2015energy}.  
Briefly, for a system with $n$ generalized coordinates $\{q_1,~q_2,..., q_n\}$, 
the PEL is the hypersurface in $(n+1)$-dimensional space defined by the potential energy of the system as a function of the 
generalized coordinates, $V({q_1,~q_2,..., q_n})$.   
The PEL formalism has been applied extensively in the past to study very diverse systems, 
including low-temperature liquids~\cite{sciortino2005potential,saika2004free,sciortino1999inherent,debenedetti2001supercooled} and glasses~\cite{sastry1998signatures,heuer2008exploring,giovambattista2016potential}, 
granular materials~\cite{liu2010jamming}, clusters of atoms~\cite{wales2015perspective,bonfanti2017methods}, proteins~\cite{wales2004energy}, and quantum liquids~\cite{giovambattista2020potential}.  
In the case of low-temperature liquids and glasses, the PEL formalism provides a simple interpretation: 
a glass is a system that, due to its low kinetic energy, is trapped within a basin of $V({q_1,~q_2,..., q_n})$
while a system in the liquid state has sufficient kinetic energy to 
move among different basins of the PEL.  
It follows that the topography of the PEL plays a fundamental role in describing the properties of low temperature
 liquids and glassy systems.

Perhaps one of the most important applications of the PEL formalism has been to provide the $P(V)$ equation of state (EOS) of liquids at low 
temperature using computer simulations, at conditions  where thermalization may be challenging~\cite{sciortino2005potential,shell2003energy,roberts1999equation,sciortino2003physics,la2002potential}. 
Such calculations are not straightforward and have been performed for only a few substances,
specifically, silica~\cite{saika2004free,saika2001fragile}, orthoterphenyl~\cite{la2002potential}, and water~\cite{roberts1999equation,sciortino2003physics,handle2018potential}.  The PEL formalism applied to the case of water has been
particularly relevant in locating the liquid-liquid critical point (LLCP) of 
 SPC/E and TIP4P/2005 water using computer simulations~\cite{sciortino2003physics,handle2018potential}.
Within the PEL formalism, one can calculate the Helmholtz free energy of the system, $F(N,V,T)$, based solely on a handful of
 topographic properties of the PEL. Specifically, $F(N,V,T)$ can be expressed in terms of (i) the average energy (depth) of the 
PEL local minima (inherent structures, IS) sampled by the system at the working conditions $(N,V,T)$, 
$E_{IS}(T)$, (ii) the curvature of the PEL about
 the sampled IS (Hessian matrix eigenvalues), and (iii) the distribution of IS energies $e_{IS}$ 
available in the PEL, $\Omega_{IS}(e_{IS})$
(these properties are accessible in computer simulations studies). 
The PEL-EOS is then calculated via $P(V)= \left( \partial F/ \partial V \right)_{N,V}$.

Most, if not all, applications of the PEL formalism to study liquids and glasses have been based on atomistic or molecular liquids composed
 of rigid molecules. The two computational studies where the PEL-EOS of water have been reported are based on the rigid SPC/E and
 TIP4P/2005 water models~\cite{sciortino2003physics,handle2018potential}.  It remains unclear whether the PEL formalism and in particular, the PEL-EOS, can be extended to the case 
of flexible molecules.  Adding molecular flexibility may lead to anharmonicities in the PEL which are difficult to deal with or, it may just lead to 
a poor agreement between the PEL-EOS and true EOS of the liquid under study.  From a fundamental point of view, classical statistical mechanics 
fails in predicting the thermodynamic properties of systems with very large vibrational 
frequencies~\cite{Habershon2011}.  For example, it is well-known that the entropy
predicted by classical statistical mechanics for an harmonic oscillator with natural frequency 
$\omega_0$ becomes negative for large $\omega_0$~\cite{aragones2007properties}. 
As shown below, the vibrational density of states of a rigid water model extends up to $\omega_c \approx 1100$~cm$^{-1}$ while 
 $\omega_c \approx 4000$~cm$^{-1}$ for flexible water models (and real water).
Within the PEL formalism, one may question how the distribution of IS (or, equivalently, the configurational entropy), 
and the energy and curvature of the IS sampled by the system differ among rigid and flexible molecular models.
In particular, one may wonder whether the PEL formalism is of any practical use to describe liquids composed
 of flexible molecules (which is the case of most, if not all, molecular force fields used in computer simulations).

In this work, we perform extensive classical molecular dynamics (MD) simulations of water using the q-TIP4P/F model~\cite{habershon2009} 
and explore the corresponding PEL. The q-TIP4P/F model is a flexible water model, where a fourth-order polynomial
 expansion of a Morse potential is used to model the OH covalent-bond potential energy, and a harmonic potential is used 
to model the HOH angle potential energy.  This model exhibits a first-order liquid-liquid phase transition (LLPT) at low temperatures that 
ends at an LLCP at positive pressures~\cite{eltareb2022evidence,eltareb2021nuclear}. The presence of an LLCP in supercooled water 
has become a common feature of computer simulations studies based on realistic water models~\cite{poole1992phase,singh2016two,biddle2017two,palmer2013liquid,palmer2014metastable,ni2016evidence}. In particular, the existence of a LLCP
in water has received remarkable strong support from recent 
experiments~\cite{kim2017maxima,kim2020experimental,pathak2021enhancement,amann2023liquid,mishima2010volume}.

The main goals of this work are (i) to test the PEL formalism for the case of the {\it flexible} q-TIP4P/F water model, (ii) to provide the PEL-EOS for this model, and (iii) to compare these results with the previous PEL studies based on the {\it rigid} SPC/E and TIP4P/2005 water models~\cite{sciortino2003physics,handle2018potential}. In particular, we test whether 
the Gaussian and Harmonic approximations of the PEL can be applied to liquids with high (intramolecular) vibrational frequencies.
 Previous studies of water and other molecular/atomistic
 systems show that the configurational entropy is related to the liquid diffusion coefficient via the Adam-Gibbs (AG) relationship~\cite{handle2018adam,chowdhary2004thermodynamics,scala2000configurational,giovambattista2003connection,adam1965temperature}. The AG relation implies that the topography of the PEL controls the dynamics of the liquid.  Accordingly, another goal of this work, is to test 
whether the AG relationship holds for q-TIP4P/F water.
Our results indicate that the EOS predicted by the PEL formalism for q-TIP4P/F water is in good agreement with 
the results from MD simulations. In particular, it provides a very good estimation of the LLCP location in the P-T plane.
Interestingly, the high vibrational frequencies in the model, due to the intramolecular interactions, lead to negative (unphysical) 
vibrational and total entropy.  Nonetheless, the configurational entropy and the thermodynamic properties of the system 
are all physically sound.   Indeed, our results with the flexible q-TIP4/F model are remarkably consistent with the previous 
PEL studies of SPC/E and TIP4P/2005 water by Sciortino and collaborators. The AG relationship, which has been validated for
 SPC/E and TIP4P/2005 water, is in good agreement with our MD results for q-TIP4P/F water. 
 
The structure of this work is as follows. In Sec.~\ref{PEL-flexible}, we present a brief overview of the PEL 
formalism, including important approximations (Gaussian and Harmonic approximations) that make the PEL formalism 
of practical use. Also included in Sec.~\ref{PEL-flexible} is the formal expression for the PEL-EOS of a flexible water model.  
The computational details are provided in Sec.~\ref{simulSec}.
The results from our classical MD simulations and PEL analysis of q-TIP4P/F water are presented in Sec.~\ref{resulSec}.
 A summary and discussions are included in Sec.~\ref{summarySec}.

\section{Potential Energy Landscape formalism for a Flexible Water Model}
\label{PEL-flexible}

We consider a flexible water model where the intermolecular interactions depend only on the position of the water
 molecules O and two H atoms. 
This is the case of the q-TIP4P/F water model (the position of the virtual site for a given water molecule is a function of the
 corresponding O and H atoms coordinates)~\cite{habershon2009}. In such cases, the canonical partition 
function for a system of $N$ molecules at temperature $T$ and volume $V$ can be written as~\cite{mcquarrie}
\begin{equation}
        Q(N,V,T) = \frac{1}{h^{3n} N!} \int_{V} dr^{3n} \int_{-\infty}^{\infty} dp^{3n} e^{-\beta H}
\label{Qeqn1}
\end{equation}
where $n=3N$ is the total number of atoms;  
$H = H(\vec{r_1}, \vec{r_2}, ...,\vec{r_n}; \vec{p_1}, \vec{p_2}, ...,\vec{p_n})$ is the Hamiltonian of the system, 
$H = \sum_{i=1}^{n} \frac{\vec{p_i}^2}{2m_i} + V(\vec{r_1}, \vec{r_2}, ..., \vec{r_n})$, and $V(\vec{r_1}, \vec{r_2}, ..., \vec{r_n})$
is the corresponding potential energy.
 Here, $m_i$ is the mass of atom $i$ and $\beta = 1/k_B T$ ($k_B$ is the Boltzmann's constant). 
The position and momentum of atom $i$ are given by $\vec{r_i}$ and $\vec{p_i}$, respectively ($i = 1, 2, .., n$). 
It follows that the configurational space of the system is a $9N$-dimensional space and that the PEL is the hypersurface in 
$(9N+1)$-dimensional space defined by the potential energy function $V(\vec{r_1}, \vec{r_2}, ..., \vec{r_n})$.

The main idea of the PEL formalism is to partition the PEL into basins~\cite{stillinger2015energy}. Each basin of 
$V(\vec{r_1}, \vec{r_2}, ..., \vec{r_n})$ is characterized by a local potential energy minimum (inherent structures, IS) 
 and the corresponding basin is defined as the set of points in $V(\vec{r_1}, \vec{r_2}, ..., \vec{r_n})$ that converge by steepest
 descent (i.e., upon potential energy minimization) to the given
 IS. It follows that each basin of the PEL can be associated with a unique 
IS characterized by an energy $e_{IS}$. It can be shown that Eq.~\ref{Qeqn1} can be re-written as~\cite{stillinger2015energy,sciortino2005potential,debenedetti2001theory} 
\begin{equation}
        Q(N,V,T) = \sum_{e_{IS}} e^{-\beta [e_{IS}- T S_{conf}(N,V,e_{IS}) + F_{vib}(N,V,T,e_{IS}) ]}
\label{Qeqn2}
\end{equation}
where the sum runs over all values of IS energies available in the PEL (if the distribution of IS energies in the PEL is continuous
then $\sum_{e_{IS}} \rightarrow \int_{e_{IS}} de_{IS}$). 
$S_{conf}(N,V,e_{IS})$  is the configurational entropy of the system and quantifies the number of IS available in the PEL with 
energy $e_{IS}$ at the given $N$ and $V$. Specifically, in Eq.~\ref{Qeqn2},
\begin{equation}
S_{conf}(N,V,e_{IS}) \equiv k_B \ln \Omega(N,V,e_{IS})
\label{SconfEqn}
\end{equation}
 where $\Omega(N,V,e_{IS})$ is the number of distinct IS available in the PEL with 
energy $e_{IS}$~\cite{sastry2001relationship,sciortino2005potential,stillinger2015energy,heuer2008exploring}.
$F_{vib}(N,V,T,e_{IS})$ is the contribution to the Helmholtz free energy arising from the vibrational motion of the system 
within the basins with IS energy $e_{IS}$~\cite{stillinger2015energy,sciortino2005potential}.
Specifically, 
\begin{equation}
F_{vib}(N,V,T,e_{IS}) \equiv -k_B T \ln{ \left(< Q_l(N,V,T) >_{l(e_{IS})} \right)}
\label{Fvib}
\end{equation}
 where  $Q_l(N,V,T)$ is the canonical partition function of the system when it is trapped in the basin of the PEL labeled by $l$, and 
$<...>_{l(e_{IS})}$ indicates averaging over all basins $l$ with IS energy $e_{IS}$ (see, e.g., Refs.~\cite{sciortino2005potential}).

{\it Gaussian and Harmonic Approximations of the PEL}.  Eq.~\ref{Qeqn2} is formally exact, equivalent to Eq.~\ref{Qeqn1}.  
However, it is of no practical use.  In order to proceed further within the PEL formalism,
 one needs to model the statistical properties of the PEL.  
The two most commonly used hypothesis in the study of liquids and glasses using the PEL formalism are (i) the Gaussian 
approximation of the PEL, which assumes that $\Omega(N,V,e_{IS})$ is a 
Gaussian distribution; and (ii) the harmonic approximation (HA),
 which assumes that the basins of the PEL have a parabolic shape about the corresponding IS~\cite{stillinger2015energy,debenedetti2001theory,sciortino2005potential,sastry2001relationship}.

(i) In the Gaussian approximation of the PEL, one assumes that 
\begin{equation}
        \Omega(N,V,e_{IS}) = \frac{e^{\alpha N}}{\sqrt{2\pi \sigma^2}} e^{-\frac{(e_{IS} - E_0)^2}{2\sigma^2}}
\label{omegaeis}
\end{equation}
where $\alpha$, $\sigma^2$, and $E_0$ are parameters that depend only on $(N,V)$. 
 The total number of IS in the PEL is given by $e^{\alpha N}$; the average IS energy and variance of $\Omega(N,V,e_{IS})$ 
are given by $E_0$ and $\sigma^2$, respectively. Using this expression in Eq.~\ref{SconfEqn}, one finds that
\begin{equation}
S_{conf}(N,V,e_{IS}) \approx k_B \left[ \alpha N - \frac{(e_{IS} - E_0)^2}{2\sigma^2} \right]
\label{Sconf2}
\end{equation}

(ii) In the HA, the vibrational free energy $F_{vib}$ can be calculated analytically (using Eq.~\ref{Fvib})~\cite{sciortino2005potential}. 
Specifically, one finds that
\begin{equation}
 F_{vib} \rightarrow F_{harm} \approx 9Nk_B T \ln(\beta A_0) + k_B T {\cal S}(N,V,e_{IS})
\label{Fharm}
\end{equation}
where ${\cal S}(N,V,e_{IS})$ is the basin shape function~\cite{sciortino2005potential}
\begin{equation}
{\cal S}(N,V,e_{IS}) = \Biggl \langle \sum_{i=1}^{9N-3} \ln \left( \frac{\hbar \omega_i(N,V,e_{IS})}{A_0} \right) \Biggr \rangle_{e_{IS}}
\label{ShapeF}
\end{equation}
and $A_0 \equiv 1$~kJ/mol is a constant that ensures that 
the argument of the $\ln(...)$  has no units. As previously found in several investigated models~\cite{sciortino2003physics,mossa2002dynamics,giovambattista2003potential}, we find that the 
shape function of the q-TIP4P/F water model is linear with $e_{IS}$ (see below),
\begin{equation}
        {\cal S}(N,V,e_{IS}) = a(N,V) + b(N,V) e_{IS}
\label{Slinear}
\end{equation}
where $a$ and $b$ are coefficients that depend only on $(N,V)$.

{\it Equilibrium.}  The key approximation in the PEL formalism is that, at a given $(N,V,T)$, the system in equilibrium only samples a
narrow range of $e_{IS}$-values~\cite{stillinger2015energy,sciortino2005potential,debenedetti2001theory}.  This is consistent with numerous computational studies~\cite{sciortino1999inherent},
 as long as the system remains in a one-phase state~\cite{altabet2016cavitation,zhou2022anomalous}. Using a saddle-point approximations~\cite{stillinger2015energy,sciortino2005potential} in Eq.~\ref{Qeqn2}, one obtains that
\begin{equation}
    Q(N,V,T) \approx  e^{-\beta [E_{IS}- T S_{conf}(N,V,E_{IS}) + F_{vib}(N,V,T,E_{IS}) ]}
\label{Qeqn3}
\end{equation}  
where the state variable $E_{IS}=E_{IS}(N,V,T)$ is the value of $e_{IS}$ that maximizes the argument in the 
exponential of Eq.~\ref{Qeqn2} (at the working conditions $(N,V,T)$). Specifically, $E_{IS}$ is the solution of the following equation,
\begin{equation}
    1 - T \left( \frac{ \partial S_{conf}(N,V,e_{IS}) }{ \partial e_{IS}} \right)_{N,V} + 
          \left( \frac{ \partial F_{vib}(N,V,T,e_{IS}) }{ \partial e_{IS}} \right)_{N,V,T} = 0
\label{Eis-eqn}
\end{equation}
In computer simulation studies, $E_{IS}(N,V,T)$ is identified with the average IS energy that the system samples at the given 
working conditions $(N,V,T)$.
Under the Gaussian and harmonic approximations of the PEL (Eqs.~\ref{omegaeis} and \ref{Fharm}), and using Eq.~\ref{Slinear}, one can solve Eq.~\ref{Eis-eqn}, resulting
in the following simple expression,
\begin{equation}
        E_{IS}(N,V,T) = E_0 - b \sigma^2 - \frac{\sigma^2}{k_B T}
\label{Eis-eqn2}
\end{equation}
where, again, $E_0$, $b$, and $\sigma^2$ all depend on $N$ and $V$.

The free energy of the system follows directly from Eq.~\ref{Qeqn3}, 
\begin{equation}
    F(N,V,T) = E_{IS}- T S_{conf}(N,V,E_{IS}) + F_{vib}(N,V,T,E_{IS}) 
\label{Feqn1}
\end{equation}
It is clear from this expression that $F_{vib}$ is, indeed, the free energy contribution arising from the exploration of the 
basins about the IS.  Similarly, $F_{IS}(N,V,T) \equiv E_{IS} - T S_{conf}$ is the 
free energy contribution to the Helmholtz free energy arising from the distribution of IS accessible to the system at the given $(N,V,T)$~\cite{stillinger2015energy}. Note that, under the Gaussian and harmonic approximations, $E_{IS}$ and $S_{conf}$ are
 given by Eqs.~\ref{Eis-eqn2} and \ref{Sconf2}, respectively; $F_{vib}$ is obtained from Eq.~\ref{Fharm} with $e_{IS} \rightarrow E_{IS}$.
Accordingly, $F(N,V,T)$ can be expressed in terms of topographic properties of the PEL, $\{\alpha,~E_0,~\sigma^2, \omega_i\}$, and $T$.

In the PEL formalism, the energy and entropy of the system are expressed as 
$E(N,V,T) \equiv  E_{IS}(N,V,T) + E_{vib}(N,V,T)$ and $S(N,V,T) \equiv S_{conf}(N,V,E_{IS}) + S_{vib}(N,V,T)$,
where $E_{vib}$ and $S_{vib}$ are the corresponding vibrational contributions, analogous to the role played by $F_{vib}$ for $F(N,V,T)$. 
Indeed, it can be shown that $F_{vib}=E_{vib} - T S_{vib}$.
Moreover, under the Gaussian and harmonic approximations of the PEL (Eqs.~\ref{omegaeis} and \ref{Fharm}), one can show that~\cite{stillinger2015energy,sciortino2005potential,debenedetti2001theory,sastry2001relationship}
\begin{equation}
E_{vib}(N,V,T) \rightarrow E_{harm}(N,V,T) = d k_b T 
\label{Evib-eqn}
\end{equation}
where $d$ is the number of degrees of freedom in the system ($d=6N$ for rigid water models; $d=9N$ for q-TIP4P/F water), and 
\begin{equation}
S_{vib}(N,V,T) \rightarrow S_{harm}(N,V,T)=  d N k_B \left[ 1 - \ln{\left( \beta \hbar A_0 \right)}  \right] - k_b {\cal S}.  
\label{Svib-eqn}
\end{equation}

{\it Anharmonic Contributions.}  
The PEL studies of SPC/E and TIP4/2005 water in Refs.~\cite{sciortino2003physics,handle2018potential} show that the HA, including  Eq.~\ref{Evib-eqn},
 do not hold for these rigid water models. 
In such cases, one needs to introduce anharmonicity corrections to the HA of the PEL~\cite{sciortino2005potential}. 
The treatment of anharmonicities in the PEL formalism 
is explained in Refs.~\cite{sciortino2005potential,handle2018potential}.  Briefly, one includes explicitly anharmonic contributions in $F_{vib}$, 
\begin{equation}
        F_{vib} = F_{harm} + F_{anharm}
\label{FharmFanh}
\end{equation}
where the first term $F_{harm}$ is given by Eq.~\ref{Fharm}; the anharmonic free energy term $F_{anharm}$
is calculated numerically.  Similarly, $E_{vib} = E_{harm} + E_{anharm}$ 
and $S_{vib}= S_{harm} + S_{anharm}$, where $E_{anharm}$ and $S_{anharm}$ are the anharmonic contribution to the energy/entropy due 
to explorations of the basins in the PEL about the corresponding IS.

In this work, we will only need the anharmonic contribution to the entropy, $S_{anharm}(N,V,T)$. 
To calculate $S_{anharm}(N,V,T)$, we follow Refs.~\cite{sciortino2005potential,handle2018potential,la2003numerical} and assume that $F_{anharm}$ is independent of 
$e_{IS}$ and depends only on $(N,V,T)$.  It can be shown~\cite{sciortino2005potential,handle2018potential} that the 
potential energy of the system can then be expressed as $U=U_{harm} + U_{anharm}$ where 
$U_{harm}=9/2 N k_B T$ is the potential energy of the system in a quadratic basin, and
 $U_{anharm}$ is the corresponding contribution due to basins anharmonicities. 
Following Refs.~{\cite{la2003numerical,handle2018potential,sciortino2005potential}, $U_{anharm}$ can be expressed as a polynomial
 in $T$ starting from the quadratic term,
\begin{equation}
        U_{anharm}(N,V,T) = c_2(V)T^2 + c_3(V)T^3
\label{UanharmEqn}
\end{equation}
where $c_2(V)$ and $c_3(V)$ are coefficients (assuming $N$ is constant). Combining Eq.~\ref{UanharmEqn}, with the relation  $dS_{anharm}/dU_{anharm} = 1/T$ (for constant $N$, $V$), the anharmonic entropy $S_{anharm}$ can be written as 
\begin{equation}
        S_{anharm}(N,V,T) = 2 c_2(V) T + \frac{3}{2}c_3(V) T^2.
\label{SanharmEqn}
\end{equation}
and $F_{anharm}$ is given by
\begin{equation}
        F_{anharm}(N,V,T) = -c_2(V) T^2  -\frac{1}{2}c_3(V) T^3.
\label{Fanharm}
\end{equation}

{\it Potential Energy Landscape Equation-of-State.} 
An analytical equation-of-state (EOS) can now be derived in the PEL formalism by using the
 thermodynamic relation $P = -(\partial F/ \partial V)_{N,T}$,
where the Helmholtz free energy $F$ (under the Gaussian and harmonic approximation, including anharmonicities) is given by
\begin{equation}
        F(N,V,T) = E_{IS}(N,V,T)- T S_{conf}(N,V,E_{IS}) + F_{harm}(N,V,T) + F_{anharm}(N,V,T).
\end{equation}
It follows that
\begin{equation}\begin{split}
        P(N,V,T)&= -\left(\frac{\partial E_{IS}(N,V,T)}{\partial V}\right)_{N,T} + T \left(\frac{\partial S_{conf}(N,V,E_{IS})}{\partial V}\right)_{N,T} \\ 
        &\qquad - \left(\frac{\partial F_{harm}(N,V,T)}{\partial V}\right)_{N,T} - \left(\frac{\partial F_{anharm}(N,V,T)}{\partial V}\right)_{N,T}
\end{split}\end{equation}

Following Refs.~\cite{sciortino2003physics,sciortino2005potential,la2002potential,la2003numerical,handle2018potential} (using Eqs.~\ref{Eis-eqn2}, \ref{Sconf2}, \ref{Fharm}, 
and \ref{Fanharm}), the PEL-EOS can then be written as (in the following, we omit the dependence of the variables on $N$)
\begin{equation}
        P(V,T) = \sum_{i=-1}^{3} {\cal P}_i(V) T^i
\label{pelEos}
\end{equation}
where ${\cal P}_i(V)$ is defined as
\begin{equation}
        {\cal P}_{-1}(V) = \frac{1}{2 k_B} \frac{d\sigma^2(V)}{dV}
\end{equation}

\begin{equation}
        {\cal P}_0(V) = -\frac{d}{dV} \left[E_0(V) - b(V) \sigma^2(V)\right]
\end{equation}

\begin{equation}
        {\cal P}_1(V) = k_B \frac{d}{dV}
        \left[N\alpha(V) - a(V) - b(V)E_0(V) + \frac{b^2(V) \sigma^2(V)}{2}\right]
\end{equation}

\begin{equation}
        {\cal P}_2(V) = \frac{dc_2(V)}{dV}
\end{equation}

\begin{equation}
        {\cal P}_3(V) = \frac{1}{2}\frac{dc_3(V)}{dV}
\end{equation}
These expressions indicate that, to calculate the PEL-EOS for $P(V)$, one only needs the PEL variables 
$\{ \alpha,~\sigma^2,~E_0;~a,~b;~c_2,~c_3\}$.  These quantities are accessible in computer simulations.

\section{Computer Simulation Details}
\label{simulSec}

We perform molecular dynamics (MD) simulations of a system composed of $N = 512$ water molecules in a cubic 
box with periodic boundary conditions. Water molecules are represented using the q-TIP4P/F model~\cite{habershon2009}. The q-TIP4P/F water model is based on the rigid TIP4P/2005~\cite{abascal2005} model, which has been used extensively to study liquid, ice, and glassy water~\cite{handle2018potential,handle2019glass,espinosa2014homogeneous,formanek2023molecular}. While the q-TIP4P/F flexible model may produce similar results to the
TIP4P/2005 rigid model, it is important to recognize that the introduction of flexibility to a water model, can have a significant impact on the corresponding phase diagram of water; see Ref.~\cite{Habershon2011,ramirez2012quasi}.
 The q-TIP4P/F water model incorporates intramolecular flexibility by modeling the O-H covalent bond
 potential with a fourth-order polynomial expansion of a Morse potential and a harmonic 
potential to model the potential energy of the HOH angle. 

The q-TIP4P/F water model has been optimized 
to be used in path-integral molecular dynamics (PIMD) simulations and it reproduces remarkably well
 the properties of liquid water~\cite{habershon2009,eltareb2021nuclear,eltareb2022evidence}, 
ice $I_h$, and LDA at $P = 0.1$~MPa~\cite{ramirez2011kinetic,ramirez2012quasi,eltareb2023}. 
Hence, one may wonder whether it is appropriate to use this model in MD simulations, as we do 
in this study. At low pressures and approximately $T > 150$~K, many of the thermodynamic, structural, 
and dynamic properties of liquid water obtained from PIMD and classical MD simulations are minor, 
if any~\cite{eltareb2021nuclear,eltareb2022evidence}, which is due to the cancellation of competing quantum effects 
in q-TIP4P/F water~\cite{habershon2009}. Differences between classical MD and PIMD simulations 
are noticeable in q-TIP4P/F liquid water at intermediate pressures and very low 
temperatures ($P = 100 - 200$~MPa and $T < 230$~K)~\cite{eltareb2022evidence}. However, even under such conditions the properties of q-TIP4P/F water reported from classical 
MD and PIMD simulations are qualitatively similar to one another, and consistent with experiments~\cite{eltareb2021nuclear,eltareb2021role,habershon2009}.

In this study we perform MD simulations at constant $N$, $V$, and $T$ over a wide range of temperatures and 
densities, $180 \leq T \leq 400$~K and $0.80 \leq \rho \leq 1.40$~g/cm$^3$; see Fig.~S1 of the Supplementary Material (SM). 
All of our MD simulations are performed using the OpenMM software package (version 7.4.0)~\cite{openMM}. 
The temperature is controlled using the stochastic (local) path-integral Langevin equation
 (PILE) thermostat~\cite{ceriotti2010}, where the thermostat collision frequency parameter is 
set to $\gamma = 0.1$~ps$^{-1}$. In our MD simulations, the time step $dt$ is set to $0.50$~fs. 
Short-range (Lennard-Jones pair potential) interactions are calculated using a cutoff $r_c = 1.0$~nm and 
the long-range electrostatic interactions are computed using the reaction-field technique~\cite{tironi1995generalized} with
 the same cutoff $r_c$. In the reaction field technique, the dielectric constant (relative permittivity) 
of the continuum medium beyond the cutoff radius $r_c$ is set to 78.3. 

In all of the MD simulations, the system is equilibrated for a time interval $t_{eq}$ followed 
by a production run of time length $t_{prod}$. The values of $t_{eq}$ and $t_{prod}$ depend on the 
state point simulated. Total MD simulation times range from $10$~ns to $2.5-4.0~\mu$s, depending on $T$ and $V$. To 
confirm that the system reaches equilibrium, we calculate the mean-square displacement (MSD) of the water 
molecules in the system as a function of time and confirm that the MD simulations satisfy the requirement
 that $t_{eq}, t_{prod} > \tau$, where $\tau$ is the time it takes for the MSD of the water molecules to reach
 $1$~nm$^2$.

{\it Inherent Structure Analysis}.  During the MD simulations, we saved a total of $25$ equally-spaced configurations 
for each state point simulated. For each of these configurations, we calculate the corresponding PEL minimum (IS) 
using the L-BFGS-B algorithm~\cite{zhu1997algorithm}. The IS energy $E_{IS}$ is obtained directly from the minimization algorithm. 
In order to calculate the curvature of the PEL basins about the corresponding IS, we obtain an analytical expression for the elements of the Hessian matrix (based on the q-TIP4P/F pair potential~\cite{habershon2009}) and evaluate it using the atoms coordinates at the IS. The Hessian matrix is composed of 
 $9N \times 9N$ ($N = 512$) elements corresponding to the second derivatives of the potential energy with respect of the coordinates of all of the atoms 
in the system. The eigenvalues of the Hessian matrix, $\{ \omega_i^2\}_{i=1,2,...,9N-3}$, are then obtained by numerically diagonalizing the Hessian matrix, which are then used to calculate the shape function defined in Eq.~\ref{ShapeF}.


{\it PEL-EOS}. The PEL-EOS (Eq.~\ref{pelEos}) depends on the PEL variables
$\{ \alpha,~\sigma^2,~E_0;~a,~b;~c_2,~c_3\}$.  These quantities are calculated as follows 
(see Refs.~\cite{sciortino2003physics,handle2018potential,sciortino2005potential,la2003numerical}).  
 For a given volume $V$ ($N$ is constant),
(i) $E_0$ and $\sigma^2$ are obtained by fitting the average IS energy as function of temperature,
 $E_{IS}(T)$, using Eq.~\ref{Eis-eqn2}; see Sec.~\ref{GausianSec}.
(ii) $a$ and $b$, are obtained by fitting the basin shape function as a function of $E_{IS}$,
 ${\cal S}(E_{IS})$, using Eq.~\ref{Slinear}; see Sec.~\ref{ShapeFun}.
(iii) $c_2$ and $c_3$ are calculated by using Eq.~\ref{UanharmEqn}; see Sec.~\ref{harmonicPEL}.
(iv) To obtain $\alpha$, we first calculate $S_{conf}(E_{IS})$ and then get $\alpha$ using Eq.~\ref{Sconf2}; see Sec.~\ref{SSconf}.  
Since $S_{conf}=S - S_{harm} - S_{anharm}$, we calculate $S$ by thermodynamic integration (see SM); 
$S_{harm}$ and $S_{anharm}$ are given by Eqs.~\ref{Svib-eqn} and \ref{SanharmEqn}.

\section{Results}
\label{resulSec}

The results are presented as follows. 
In Secs.~\ref{Water-Spectra} and \ref{llcp}, we discuss basic properties of q-TIP4P/F water obtained
 from MD simulations. Specifically, we study (A) the vibrational density of states calculated at the IS, 
and (B) the phase diagram of this water model, which exhibits 
a LLPT and LLCP at low temperatures. 
In Sec.~\ref{gaussianPEL} we show that the Gaussian approximation successfully applies to the PEL of q-TIP4P/F water (Sec.~\ref{PEL-flexible}).
Secs.~\ref{harmonicPEL} and \ref{ShapeFun} focus on the harmonic approximation of the PEL of q-TIP4P/F water, including the calculation of the basin shape function and corrections due to the basins anharmonicities. The calculations of  $S(T)$ and $S_{conf}(T)$ are discussed in Sec.~\ref{SSconf}. 
The EOS of q-TIP4P/F water derived from the parameters of the PEL is presented in Sec.~\ref{phase-diagram}. 
Lastly, in Sec.~\ref{adam-gibbs}, we confirm the validity of the Adam-Gibbs for 
q-TIP4P/F water.

\subsection{Vibrational Spectra of q-TIP4P/F water from the IS}
\label{Water-Spectra}
Fig.~\ref{VDOSfig} shows the vibrational density of states (VDOS) of q-TIP4P/F water
 obtained by diagonalization of the Hessian matrix at the IS. The effect of density at a fix temperature ($T = 220$~K) is shown
in Figs.~\ref{VDOSfig}(a) and \ref{VDOSfig}(b); the effect of temperature at a fix density ($\rho = 1.00$~g/cm$^3$)
is shown in Figs.~\ref{VDOSfig}(c) and \ref{VDOSfig}(d).
For clarity, the VDOS in Fig.~\ref{VDOSfig} are split into low ($\omega < 1400$~cm$^{-1}$)  
and high frequency ($\omega > 1400$~cm$^{-1}$) ranges, highlighting the 
(i) translational ($\omega < 400$~cm$^{-1}$), 
(ii) librational ($400 < \omega <1200$~cm$^{-1}$),
(iii) HOH bending ($\omega \approx 1600$~cm$^{-1}$), and 
(iv) O-H stretching bands ($\omega \approx 3400 - 3600$~cm$^{-1}$).
The effects of increasing the temperature (at constant density), and the density (at constant temperature) are qualitatively similar.
Specifically, increasing the temperature/density tends to reduce the magnitude of the VDOS at the peaks corresponding to all vibrational modes (i)-(iv).
The only exception to this are the translational modes centered at $\omega \approx 50$~cm$^{-1}$, which remains unchanged upon heating [Fig.~\ref{VDOSfig}(c)].
A slight shift in the VDOS peaks corresponding to vibrational modes (i), (ii) and (iv) are also found upon increasing temperature/density.
For comparison, also included in Figs.~\ref{VDOSfig}(b) and \ref{VDOSfig}(d) are the vibrational modes of
a single water molecule (black dashed-lines). For a single water molecule there are three distinct frequencies, 
corresponding to the bending ($\omega \approx 1600$~cm$^{-1}$), and symmetric/asymmetric O-H stretching modes ($\omega \approx 3950$~cm$^{-1}$ and
$\omega \approx 4050$~cm$^{-1}$).  The bending and stretching bands in q-TIP4P/F water are slightly shifted relative to the 
corresponding frequencies for an isolated water molecule

We stress that the VDOS of rigid water models extend up to $\omega \approx 1200$~cm$^{-1}$. Such VDOS are similar to the VDOS 
shown in Fig.~\ref{VDOSfig}(a) and \ref{VDOSfig}(c). Relevant to this work,  
the changes in the VDOS of q-TIP4P/F water due to temperature and density, shown in 
Figs.~\ref{VDOSfig}(a) and \ref{VDOSfig}(c), are fully consistent with the corresponding effects observed in the VDOS of TIP4P/2005 
water reported in Ref.~\cite{handle2018potential}.


\begin{figure}[!htb]
\centering{
		\includegraphics[width=7.0cm]{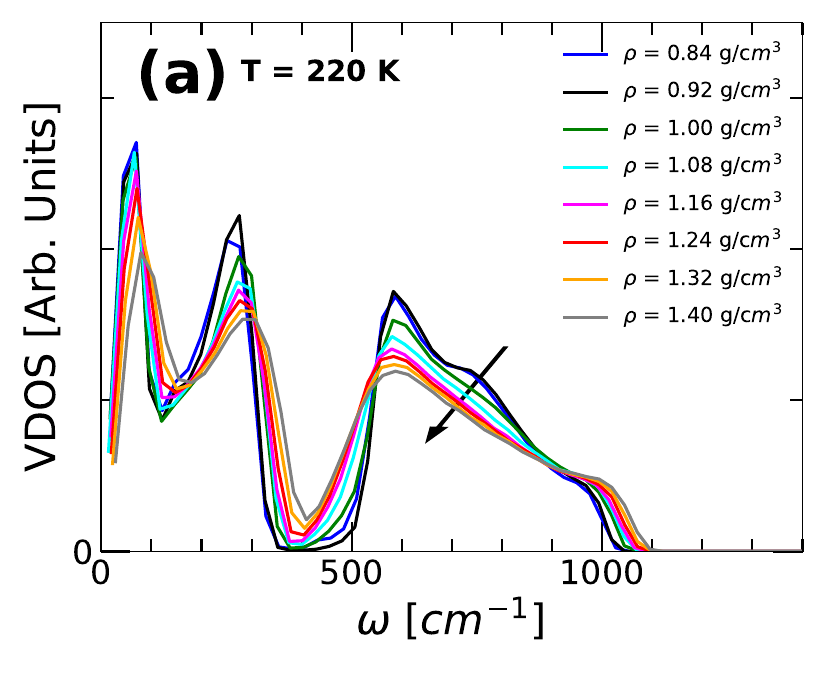}
		\includegraphics[width=7.0cm]{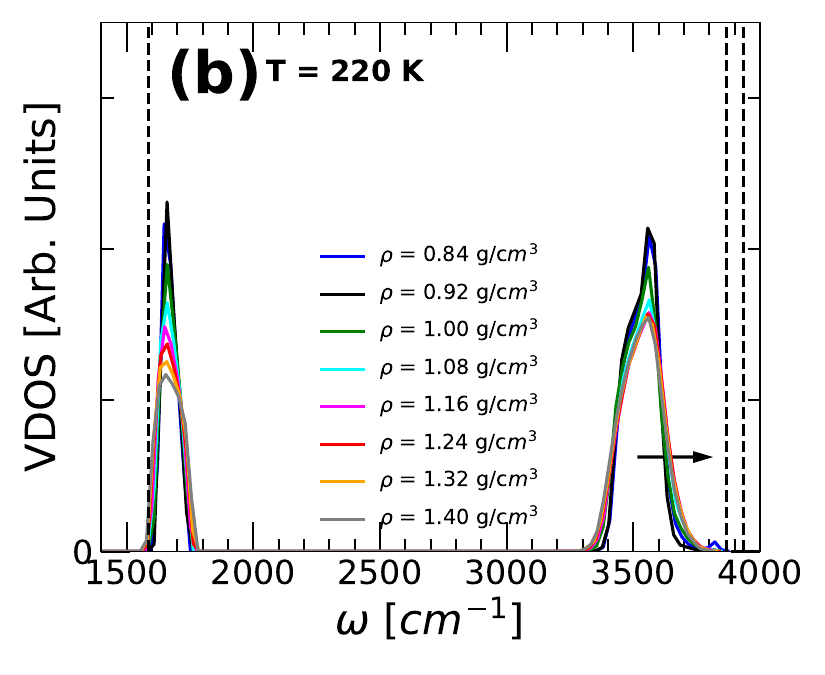}
		\includegraphics[width=7.0cm]{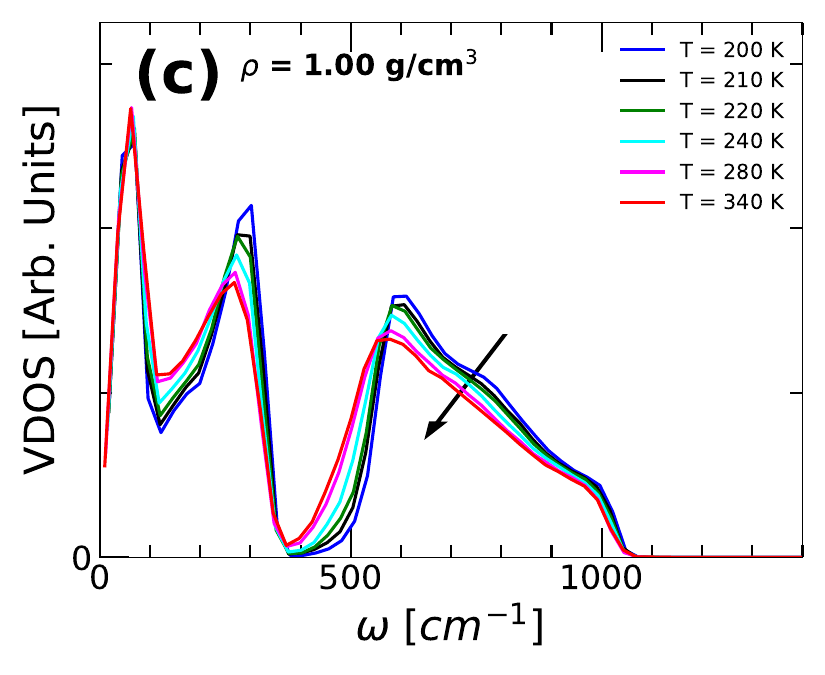}
		\includegraphics[width=7.0cm]{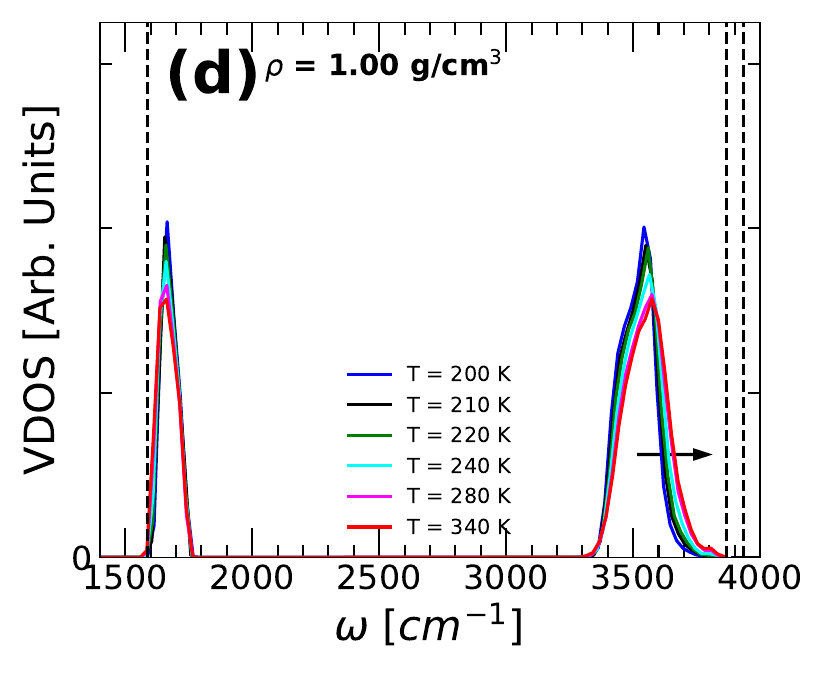}
}
\caption{Vibrational density of states of q-TIP4P/F water obtained by diagonalizing the Hessian matrix at the IS sampled by the system.
(a) VDOS as a function of density at $T = 220$~K for frequencies $\omega < 1400$~cm$^{-1}$, covering the 
librational mode frequencies at approximately $\omega>400$~cm$^{-1}$ and the translational mode frequencies 
at approximately $\omega<400$~cm$^{-1}$. The VDOS maxima corresponding to both the librational and translational vibrational modes 
 of the system decrease with increasing density. 
(b) Same as (a) for frequencies $1400 \leq \omega \leq 4000$~cm$^{-1}$, corresponding to the bending ($\omega<1600$~cm$^{-1}$) 
and stretching vibrational modes ($\omega<3500$~cm$^{-1}$). Increasing the density of the system also 
reduce the VDOS maxima associated to the bending and stretching modes.
(c)(d) Same as (a)(b) for the VDOS as a function of temperature, at density $\rho = 1.00$~g/cm$^3$.
Increasing the temperature has a similar effect on the VDOS as increasing the density (see arrows).
For comparison, included in (b) and (d) are the frequencies of an isolated water molecule associated to
the HOH bending ($\omega \approx 1600$~cm$^{-1}$) and OH stretching modes ($\omega \approx 3950$~cm$^{-1}$ for the symmetric mode, and
$\omega \approx 4050$~cm$^{-1}$ for the asymmetric mode). 
The changes in the VDOS with temperature and density at $\omega < 1400$~cm$^{-1}$ in (a) and (c) are fully consistent with the corresponding 
results obtained in MD simulations using rigid water models~\cite{handle2018potential}.
}
\label{VDOSfig}
\end{figure}
\clearpage

\subsection{Liquid-liquid phase transition}
\label{llcp}

Fig.~\ref{PTfig}(a) shows the $P-T$-phase diagram of q-TIP4P/F water obtained from our MD simulations (using the reaction-field technique to treat the electrostatic interactions). Included in Fig.~\ref{PTfig}(a)
are the LLCP, line of isothermal compressibility maxima ($\kappa_{T}^{max}$-line), line of 
density maxima ($\rho^{max}$-line), and the line of diffusivity maxima ($D^{max}$-line)~(see, e.g., Refs.~\cite{debenedetti2003supercooled,poole2005density,eltareb2022evidence,giovambattista2013liquid}). 
Also included is the liquid-vapor spinodal line.
  The black circles/lines in Fig.~\ref{PTfig}(a) are selected isochores. Consistent with thermodynamics~\cite{debenedetti1996metastable}, 
the isochores for volumes close to the LLCP volume intersect at the LLCP (red star). 
We note that the LLCP in Fig.~\ref{PTfig}(a) is determined from the inflection point in the $P(V)$-isotherms; see Fig.~\ref{PTfig}(b).
Indeed, the $P(V)$ isotherms shown in Fig.~2(b) exhibit a region of instability where $\left(\partial P/ \partial V\right)_{N,V} > 0$ at $T<190$~K indicating that 
the system is separating into two distinct liquid phases, low-density and high-density liquid (LDL and HDL). 
Unfortunately, below $T < 200$~K and for approximately $1.05< V <1.15$~cm$^3$/g, we could not perform MD simulations 
due to the very slow relaxation of the system ($>10$~$\mu$s). Crystallization is not observed in most of the state points explored, except in a few runs performed at $T = 190-200$~K, below the LLCP, and at $\rho = 1.00$~g/cm$^3$; see SM.

The potential energy of the system $U(V)$ along the isotherms shown in Fig.~\ref{PTfig}(b) are shown in Fig.~\ref{PTfig}(c). 
$U(V)$ exhibits a concave region (i.e., $(\partial^2 U/ \partial V^2)_{N,T} < 0$) at $V = 0.85-1.00$~cm$^3$/g, particularly at low temperatures.
Since the Helmholtz free energy of the system is $F = E - TS$, a concavity in $U(V)$ 
 can lead to a concavity in $F(V)$ at low temperatures, which is a signature of a first-order phase transition.
Therefore, the concavity in $U(V)$ is fully consistent with the existence of a LLPT/LLCP 
in q-TIP4P/F water~\cite{callen2006thermodynamics}.
The liquid-vapor spinodal line (orange line) shown in Fig.~\ref{PTfig}(a) is 
determined from the minimum in the $P(V)$-isotherms at $V > 1.1$~g/cm$^3$ in Fig.~\ref{PTfig}(b).


Overall, the phase diagram in Fig.~\ref{PTfig}(a) and the $P(V)$- and $U(V)$-isotherms shown in Fig.~\ref{PTfig}(b) and \ref{PTfig}(c) 
are fully consistent with previous computational studies using different water models, including 
ST2~\cite{poole1992phase,poole2005density,poole2013free,liu2012liquid,holten2014two}}, TIP4P/2005 and TIP4P/ice~\cite{gonzalez2016comprehensive,abascal2010widom,debenedetti2020second}. 
We also note that the results presented here, based on the q-TIP4P/F model using the reaction field technique, are very similar to
the results obtained in Ref.~\cite{eltareb2022evidence} using the same water model but with the Particle Mesh Ewald (PME) technique. 
Using the reaction field technique shifts the LLCP location from $(P_c=203$~MPa,~$T_c=175$~K, $\rho_c=1.03$~g/cm$^3)$ [PME, {\it estimated} in Ref.~\cite{eltareb2022evidence})]
to $(P_c=150$~MPa,~$T_c=190$~K, $\rho_c=1.04$~g/cm$^3)$ [reaction field, this work].

\begin{figure}[!htb]
\centering{
	\includegraphics[width=7.0cm]{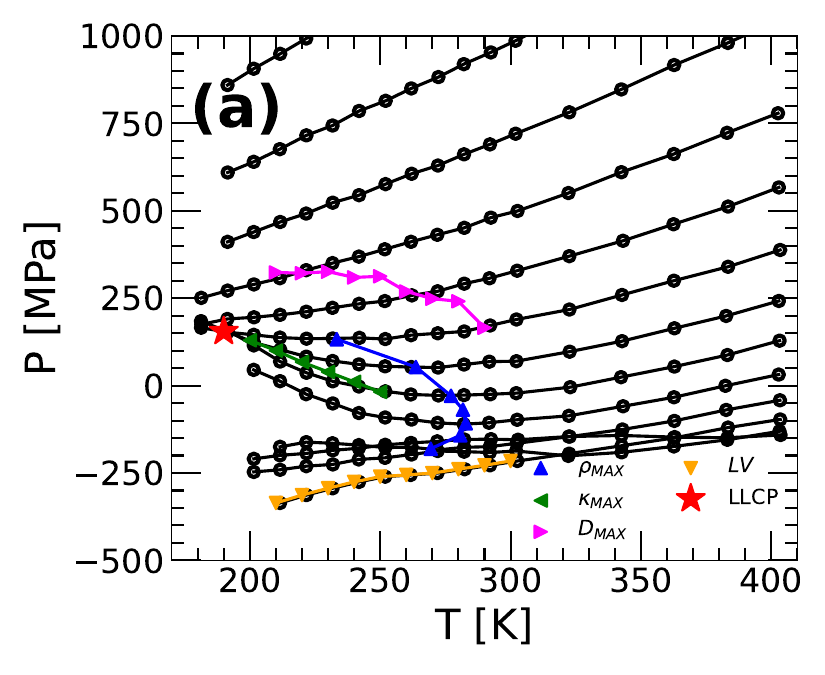}
	\includegraphics[width=7.0cm]{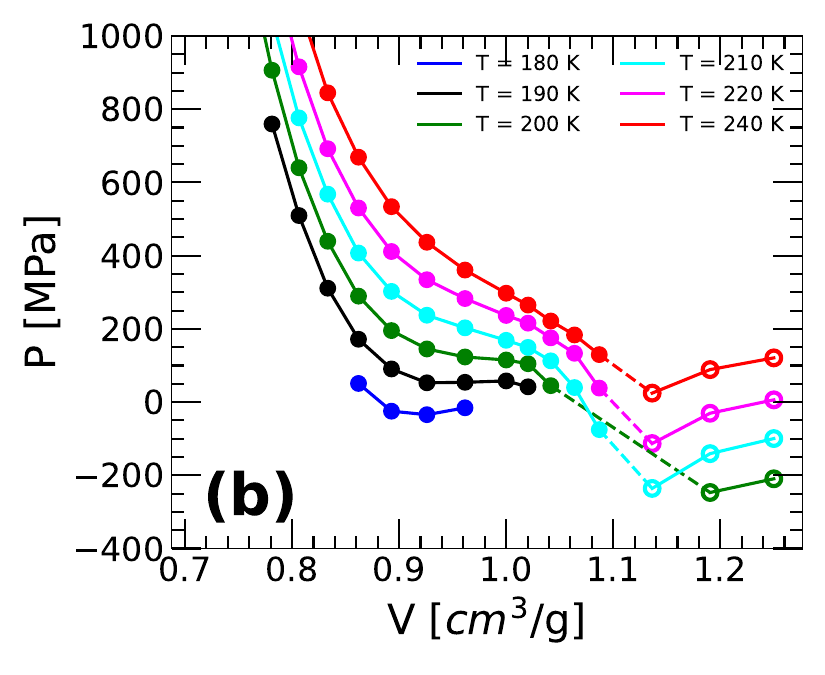}
	\includegraphics[width=7.0cm]{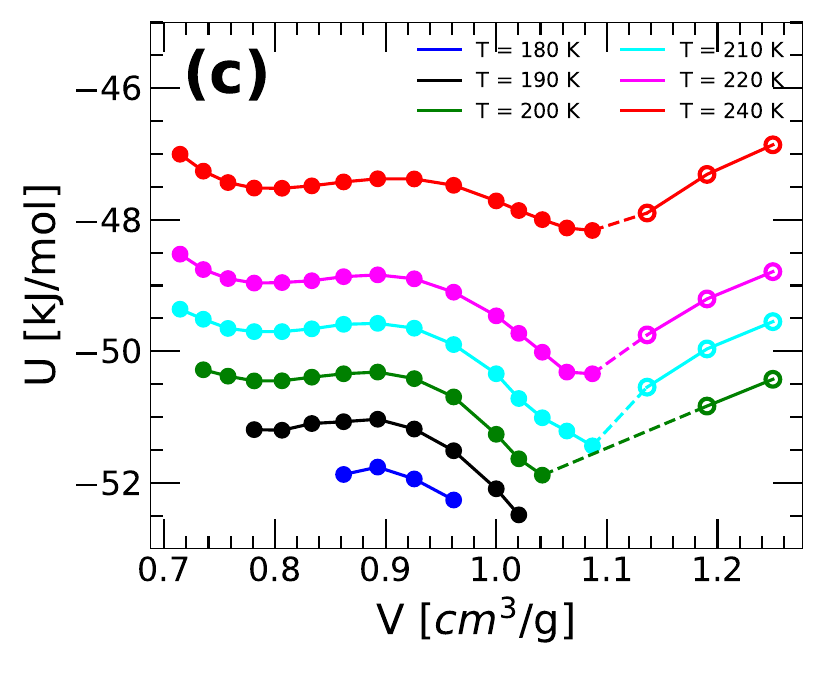}
}
\caption{(a) $P-T$-phase diagram of q-TIP4P/F water obtained from classical MD simulations. MD simulations are performed 
down to $T=180$~K, just below the LLCP temperature (red star). Black lines/circles are the isochores 
for (top-to-bottom) $\rho=0.80$ to $1.40$~g/cm$^3$ in steps of $0.04$~g/cm$^3$ -- isochores for $\rho \approx 1.00-1.04$~g/cm$^3$
 intersect one another at the LLCP.
The liquid-vapor spinodal line is indicated by the orange down-triangles. The green left-triangles and blue up-triangles represent, respectively, 
the lines of maxima in the isothermal compressibility and density; magenta right-triangles indicate the line of maxima in the 
diffusion coefficient.
(b) $P(V)$-isotherms of q-TIP4P/F water for selected temperatures.  Solid and empty circles correspond to liquid and vapor states, respectively. 
Equilibration was not reached at $V \approx 1.04-1.14$~cm$^3$/g ($\rho \approx 0.88-0.96$~g/cm$^3$) due to slow relaxation times (dashed lines). 
Isotherms are shifted for clarity by (top-to-bottom) $\Delta P = 300,~200,~100,~0,~-100,~-200$~MPa. 
At low temperatures, the $P(V)$-isotherms develop an inflection point consistent with the existence of a LLPT in q-TIP4P/F water at
 $(P_c=150$~MPa,~$T_c=190$~K, $\rho_c=1.04$~g/cm$^3)$.
(c) Potential energy ($U(V)$) as a function of volume for q-TIP4P/F water along the isotherms shown in (b). Consistent with the existence of a LLCP,
the $U(V)$-isotherms exhibit a concavity region at intermediate volumes.
}
\label{PTfig} 
\end{figure}
\clearpage

\subsection{Gaussian Approximation}
\label{GausianSec}

\begin{figure}[!htb]
\centering{
	\includegraphics[width=8.0cm]{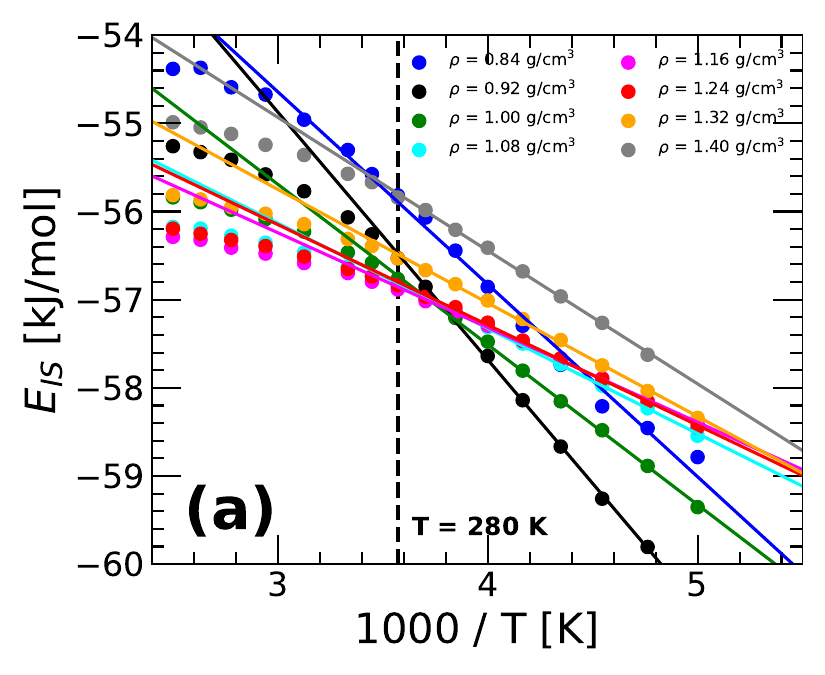}
	\includegraphics[width=8.0cm]{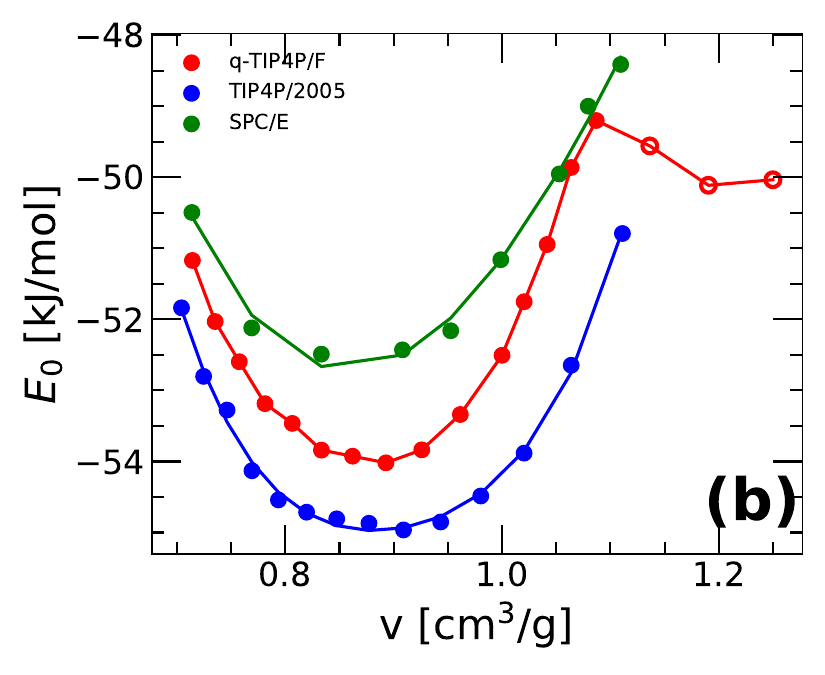}
	\includegraphics[width=8.0cm]{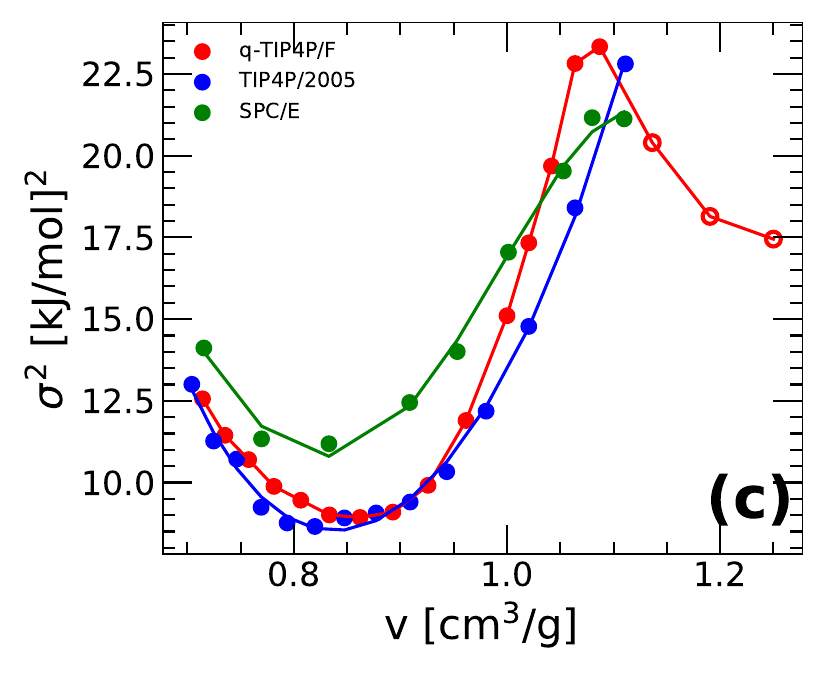}
}
\caption{(a) Inherent structure energy $E_{IS}$ of q-TIP4P/F water as a function of the inverse temperature for selected isochores. 
Below $T \leq 280$~K (vertical dashed-line), $E_{IS}(T)$ is a linear function of $1/T$ (solid lines), 
consistent with the Gaussian approximation of the PEL, Eq.~\ref{Eis-eqn2}. 
(b)(c) PEL parameters $E_0$ and $\sigma^2$ as function of volume obtained from the linear fits in (a) using Eq.~\ref{Eis-eqn2}.
For comparison, also included in (b) and (c) are the $E_0(V)$ and $\sigma^2(V)$ values reported for TIP4P/2005 (blue circles) 
and SPC/E (green circles) in Refs.~\cite{handle2018potential,sciortino2003physics}. The values of 
$E_0(V)$ and $\sigma^2(V)$ for q-TIP4P/F, TIP4P/2005, and SPC/E water are very similar to one another and
show the same qualitative dependence on $V$. Solid and empty circles correspond to liquid and vapor states.
}
\label{Eis-fig}
\end{figure}

In this section, we show that the Gaussian approximation of the PEL is consistent with the MD simulations results of q-TIP4P/F water.
To do this, we include in Fig.~\ref{Eis-fig}(a) the average IS energy $E_{IS}(T)$ as a function of the inverse temperature $1/T$
for isochores above and below the critical isochore for q-TIP4P/F ($\rho_c \approx 1.04$~g/cm$^3$).
The main point of Fig.~\ref{Eis-fig}(a) is that, at all densities considered, $E_{IS}(T) \propto 1/T$ for $T < 280$~K, 
consistent with the prediction of the Gaussian approximation (Eq.~\ref{Eis-eqn2}). 
We note that at very high temperature ($T>400$~K),
 $E_{IS}(T)$ reaches a constant value, in the so-called PEL-independent 
regime~\cite{starr2001thermodynamic,sastry1998signatures}; see Fig.~S3 of the SM. In this regime, the properties of the system are not sensitive to the topography 
of the PEL because the system has a large kinetic energy and hence, it can freely explore the PEL. 
Instead, at low temperature, the thermodynamic properties of the system are strongly correlated with the topography of the PEL, 
in the so-called PEL-influenced regime~\cite{sastry1998signatures}, where $E_{IS}(T)$ varies non-linearly with 
respect to $T$.  The crossover temperature between the PEL-independent and PEL-influenced regime occurs at the so-called onset 
temperature $T_0$~\cite{giovambattista2020potential}. For the case of q-TIP4P/F the onset temperature is $T_0 \approx  280-330$~K, 
which is larger than the melting temperature for this water model $T_M \approx 260$~K~(using the Ewald summation technique~\cite{habershon2009}). 

The parameters $E_0(V)$ and $\sigma^2(V)$ for q-TIP4P/F water are evaluated from Fig.~\ref{Eis-fig}(a) 
by interpolating the MD simulations data (circles) at low temperatures using a straight line (Eq.~\ref{Eis-eqn2}).  
$E_0(V)$ and $\sigma^2(V)$ are shown in Fig.~\ref{Eis-fig}(b) and \ref{Eis-fig}(c) (red circles). For comparison,
 we also include the values of $E_0(V)$ and $\sigma^2(V)$ for TIP4P/2005 (blue circles) and SPC/E (green circles) water
reported in Refs.~\cite{sciortino2003physics,handle2018potential}. The values and volume-dependence of $E_0(V)$ and $\sigma^2(V)$ for the q-TIP4P/F water
 are qualitatively similar to the corresponding values obtained for the rigid water SPC/E and TIP4P/2005 models.
Accordingly, irrespective of whether the model is rigid (SPC/E and TIP4P/2005) or flexible (q-TIP4P/F), 
a Gaussian description is a good approximation for the PEL of water at low temperatures. 
Interestingly, for all of these three water models, 
a minimum in $E_0(V)$ and $\sigma^2(V)$ occurs at $V \approx 0.85-0.90$~cm$^3$/g.
The minimum in $\sigma^2(V)$ is particularly important.  It has been shown that for liquids with a PEL that is Gaussian and harmonic, 
a minimum in $\sigma^2(V)$ implies that the liquid has a density anomaly (at densities where $d\sigma^2/dV > 0$)~\cite{handle2018potential,sciortino2003physics}.
In addition, the minimum in $\sigma^2(V)$ also implies that, for a Gaussian and harmonic PEL, the corresponding liquid exhibits a LLCP~\cite{sciortino2003physics}.
Hence, our results are fully consistent with the previous studies of Sciortino and collaborators 
based on the rigid SPC/E and TIP4P/2005 water models~\cite{sciortino2003physics,handle2018potential} and with Ref.~\cite{la2002potential}.


\label{gaussianPEL}
\clearpage

\subsection{Harmonic Approximation}
\label{harmonicPEL}

In this section, we test whether the HA of the PEL applies to the q-TIP4P/F water model. 
Briefly, we find that the HA of the PEL does not hold for q-TIP4P/F water at all temperatures and volumes studied, 
similar to the case of rigid water models~\cite{starr2001thermodynamic,handle2018potential,sciortino2003physics}. This implies that the PEL basins of q-TIP4P/F water contain 
significant anharmonicities that need to be taken into account when applying the PEL formalism (see Sec.~\ref{PEL-flexible}).   

\begin{figure}[!htb]
\centering{
	\includegraphics[width=8.0cm]{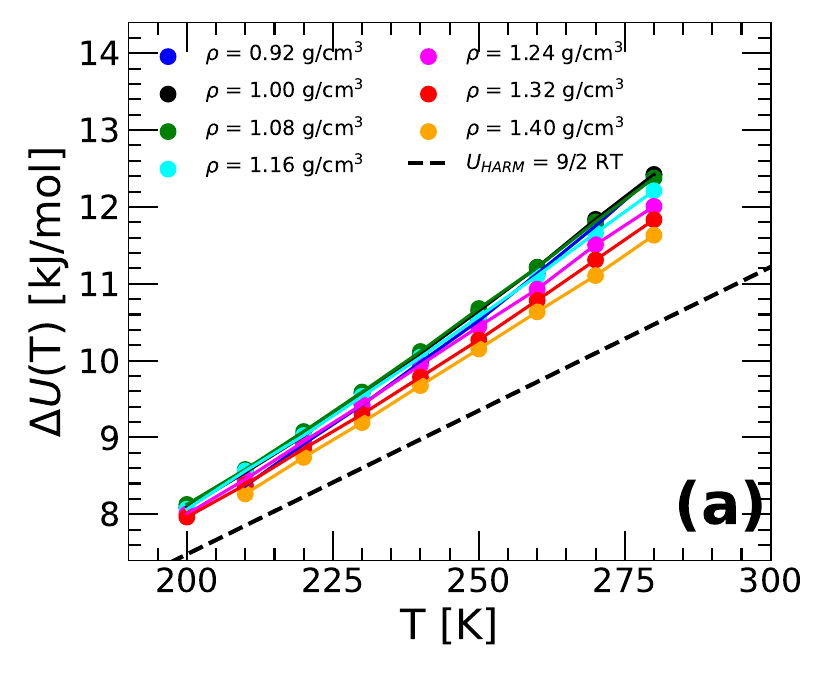}
	\includegraphics[width=8.0cm]{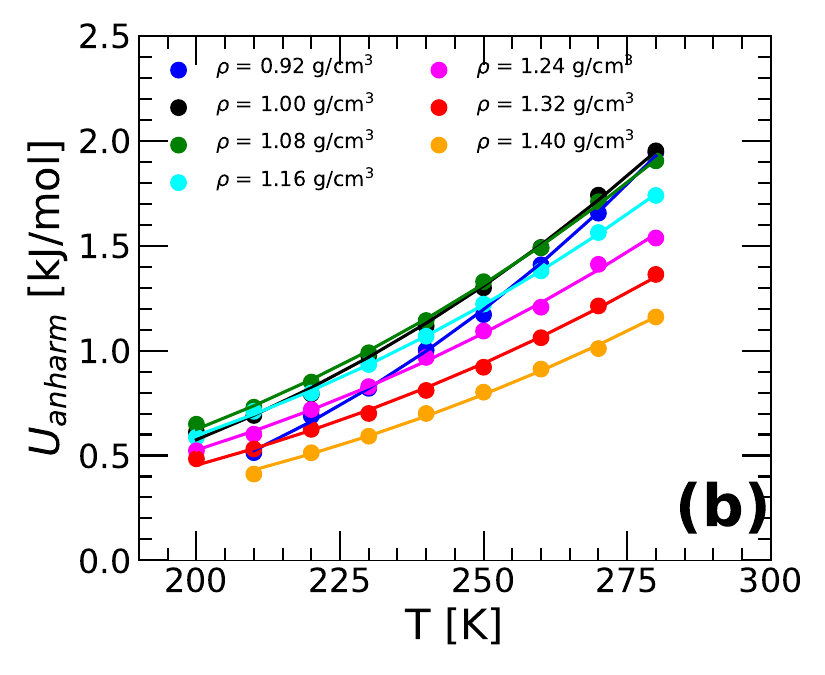}
	\includegraphics[width=8.0cm]{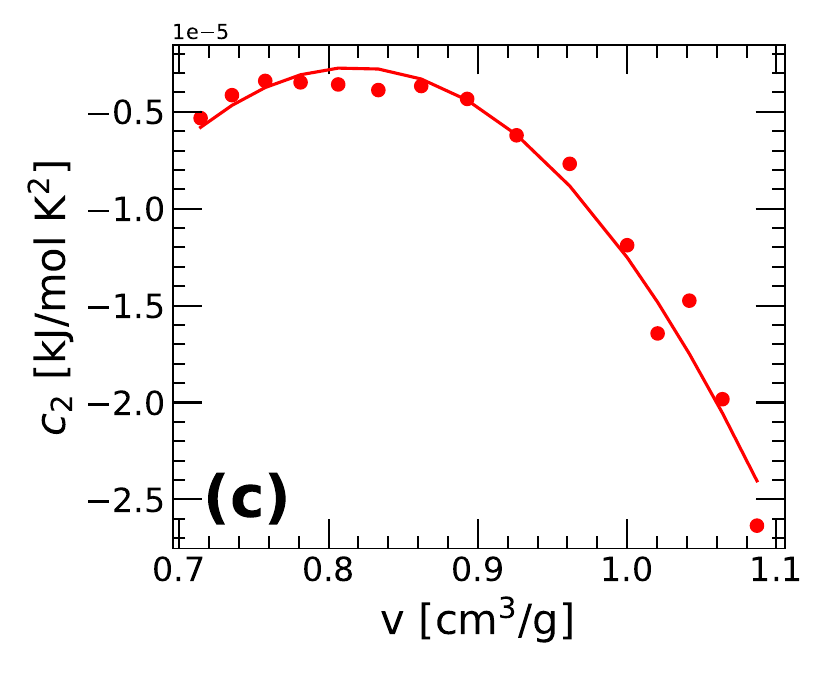}
	\includegraphics[width=8.0cm]{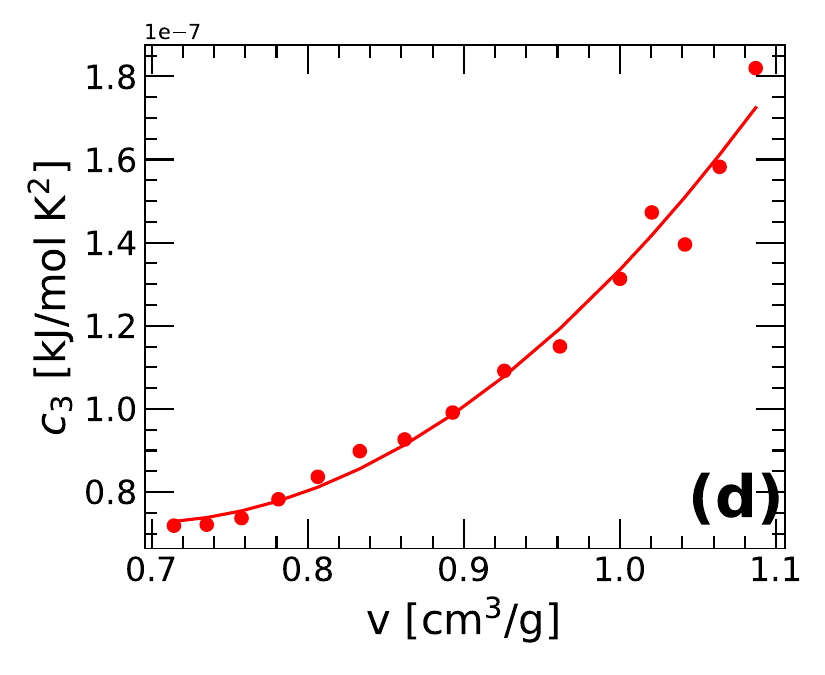}
}
\caption{(a) Potential energy of q-TIP4P/F water $U(T)$ minus the IS energy $E_{IS}$ along different isochores
 ($\Delta U(T)= U(T) -E_{IS}(T)$; circles).  The prediction from the HA approximation of the PEL is indicated by the dashed-line
($\Delta U(T) = U_{harm} = \frac{9}{2} N k_B T$).  For all the isochores studied, the HA approximation does not hold for q-TIP4P/F water.
(b) Contribution to the potential energy of the system due to the basin anharmonicities, $U_{anharm}(T)$.
 The lines are the fits to the MD data obtained using Eq.~\ref{UanharmEqn}. 
(c)(d) Fitting parameters $c_2(V)$ and $c_3(V)$ defined in Eq.~\ref{UanharmEqn} and obtained from the fittings in (b).}
\label{Uharm-fig}
\end{figure}

To test whether the HA holds for q-TIP4P/F water, we focus on the potential energy of the system, $U(N,V,T)$.
Within the HA of the PEL, at given $N$ and $V$, $U(T)= E_{IS}(T) + U_{harm}(T)$ with $U_{harm}= \frac{9}{2} N k_B T$
 (see, e.g., Refs.~\cite{sciortino2005potential,handle2018potential,starr2001thermodynamic}). As shown in Fig.~\ref{Uharm-fig}(a), this expression does not hold for q-TIP4P/F water for
 any of the isochores studied.  

The contribution to the potential energy due to the anharmonicities in the basins of the PEL is given by
\begin{equation}
U_{anharm}(T)= U(T) - E_{IS}(T) - \frac{9}{2} N k_B T
\label{Uanharm2}
\end{equation}
Fig.~\ref{Uharm-fig}(b) shows $U_{anharm}(T)$ obtained from Fig.~\ref{Uharm-fig}(a) (circles) as well as 
the corresponding fitting curves based on Eq.~\ref{UanharmEqn}.
The excellent agreement between the MD simulation data and Eq.~\ref{UanharmEqn} indicates that one can treat the 
anharmonic corrections to the PEL using Eqs.~\ref{UanharmEqn}-\ref{Fanharm}.
 $c_2(V)$ and $c_3(V)$ are shown in Figs.~\ref{Uharm-fig}(c) and \ref{Uharm-fig}(d), respectively.

\clearpage

\subsection{Shape Function}
\label{ShapeFun}

\begin{figure}[!htb]
\centering{
	\includegraphics[width=8.0cm]{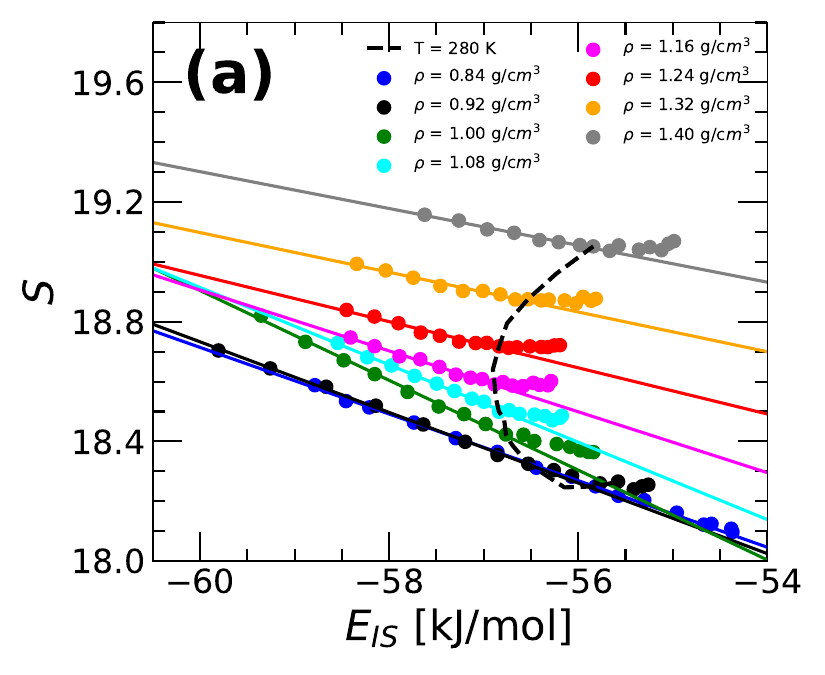}
	\includegraphics[width=8.0cm]{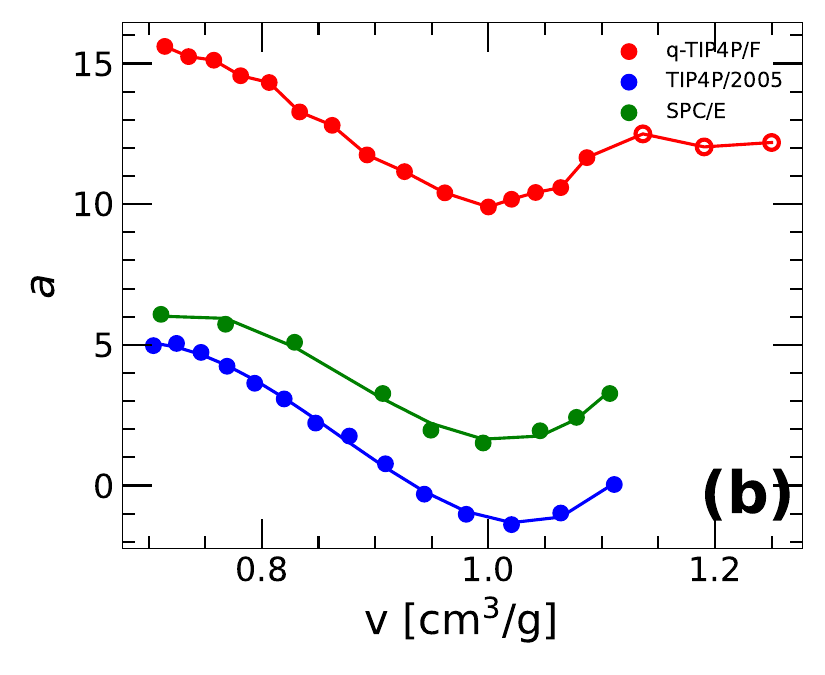}
	\includegraphics[width=8.0cm]{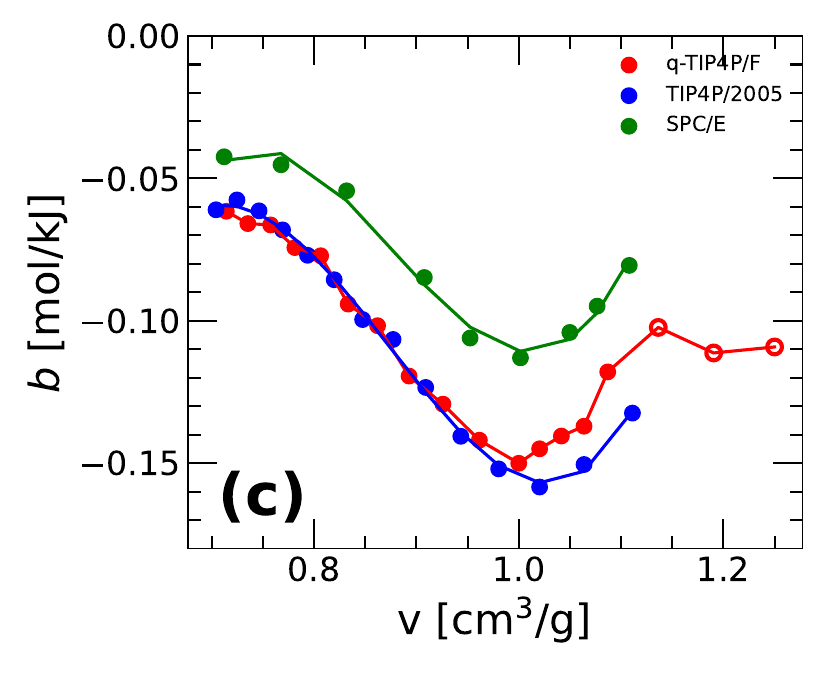}
}

\caption{(a) Basin shape function ${\cal S}$ as a function of the IS energy $E_{IS}$ 
for q-TIP4P/F water along selected isochores.
Circles are results  from MD simulations; lines are the linear fit of the MD data for $T \leq 280$~K (the dashed line corresponds to the ${\cal S}-E_{IS}$ data points at $T = 280$~K).
In this temperature-range, ${\cal S} \propto E_{IS}$ consistent with Eq.~~\ref{Slinear}. 
(b)(c) The PEL parameters $a(V)$ and $b(V)$ defined in Eq.~\ref{Slinear} and resulting from the linear fittings in (a).
For comparison, we also include the values of $a(V)$ and $b(V)$ for TIP4P/2005 (blue circles) 
and SPC/E (green circles) reported in Refs.~\cite{handle2018potential,sciortino2003physics}. 
 $a(V)$ and $b(V)$ exhibit qualitatively similar $V$-dependence, irrespective of whether 
the water model is flexible (q-TIP4P/F) or rigid (TIP4P/2005 and SPC/E); the value of $a(V)$ are larger in the case of q-TIP4P/F water. }
\label{ShapeF-fig}
\end{figure}

Next, we show that the basin shape function ${\cal S}(N,V,T)$ obeys Eq.~\ref{Slinear} for the case of q-TIP4P/F water.
We obtain ${\cal S}(N,V,T)$ by diagonalizing the Hessian matrix of the system evaluated at the IS sampled
during the MD simulations, and using Eq.~\ref{ShapeF}.    
Fig.~\ref{ShapeF-fig}(a) shows the ${\cal S}(T)$ of q-TIP4P/F water along selective isochores. It follows that, 
for approximately $T \leq 280$~K,  ${\cal S}(T)$ is a linear function of $E_{IS}$, consistent with Eq.~\ref{Slinear}.
We note that the temperature range $T \leq 280$~K, at which Eq.~\ref{Slinear} holds, is also the range of temperatures
where $E_{IS} \propto 1/T$ [Fig.~3(a)], i.e., where the Gaussian approximation of the PEL holds for q-TIP4P/F water.

From the linear fittings in Fig.~\ref{ShapeF-fig}(a), we extract the fitting parameters $a(V)$ and $b(V)$ defined in
Eq.~\ref{Slinear}. The parameters $a(V)$ and $b(V)$ for q-TIP4P/F water are shown in Fig.~\ref{ShapeF-fig}(b) and 
\ref{ShapeF-fig}(c) (red circles) together with the corresponding values for
 the TIP4P/2005 (blue circles) and SPC/E (green circles) water models reported in Refs.~\cite{handle2018potential,sciortino2003physics}. 
As for the cases of $E_0(V)$ and $\sigma^2(V)$, we also find that the $V$-dependence of
 $a(V)$ and $b(V)$ are similar for all of three water  models. While the values of $b(V)$ are quantitatively very 
similar in all these models, the values of $a(V)$ in q-TIP4P/F water
are much larger than those reported for SPC/E and TIP4P/2005 water models.
This is because the frequencies $\{ \omega_i \}$ of q-TIP4P/F water include the OH stretching and HOH angle bending bands,
 which contribute significantly to the shape function (see Eq.~\ref{ShapeF}.).
Nonetheless, the slopes of $a(V)$ and $b(V)$ do not appreciably change among the q-TIP4P/F, TIP4P/2005, and SPC/E models.

\clearpage

\subsection{Entropy and Configurational Entropy} 
\label{SSconf}

In this section we calculate the (i) $S$ and (ii) $S_{conf}$ of q-TIP4P/F water.  By doing so, we also obtain 
the PEL variable $\alpha(V)$ which is needed to calculate the PEL-EOS. Our calculations 
follow the same procedure employed in Ref.~\cite{handle2018potential}.

\begin{figure}[!htb]
\centering{
	\includegraphics[width=8.0cm]{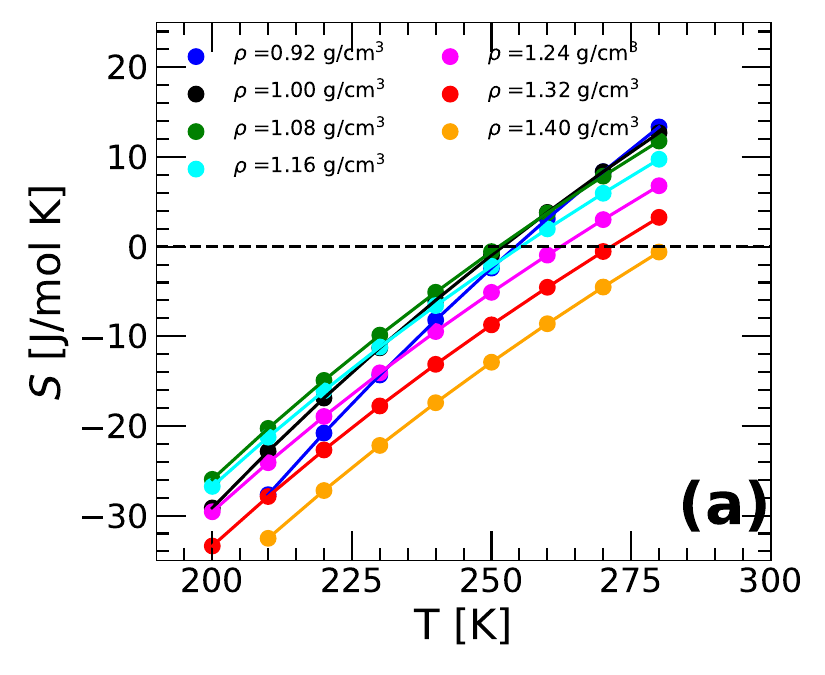}
	\includegraphics[width=8.0cm]{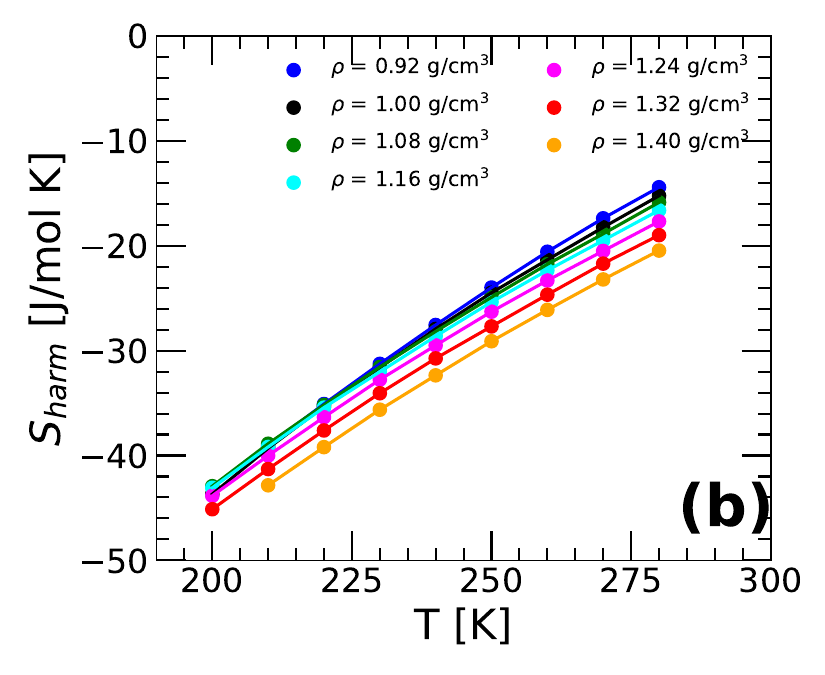}
	\includegraphics[width=8.0cm]{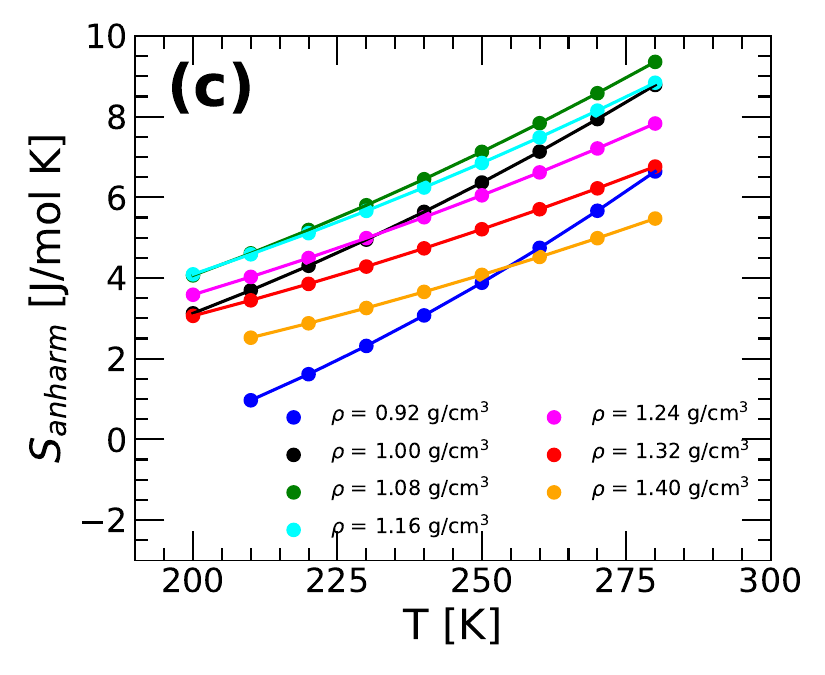}
	\includegraphics[width=8.0cm]{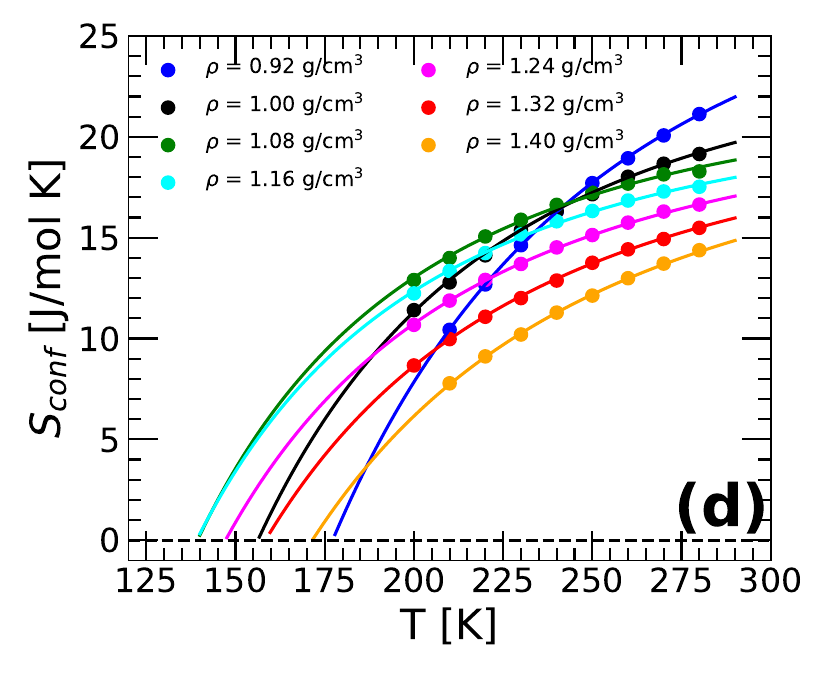}
}
\caption{(a) Entropy $S(T)$ of q-TIP4P/F water as a function of temperature obtained from the MD simulations for selected isochores via thermodynamic integration (see SM).
$S(T)$ is unphysically negative at low temperatures, a consequence of the high-frequency bending and stretching 
modes of this flexible water model (see text)~\cite{Habershon2011}.
(b)(c) Contribution to the total entropy from the harmonic and anharmonic part of the vibrational entropy, $S_{harm}(T)$ and $S_{anharm}(T)$.
$S_{harm}$ and $S_{anharm}$ are  calculated using Eqs.~\ref{Svib-eqn} and \ref{SanharmEqn}, respectively. 
(d) Configurational entropy $S_{conf}(T) = S(T) - S_{harm}(T) - S_{anharm}(T)$ obtained from (a)-(c). Lines are the best fits predicted by the
Gaussian approximations of the PEL, Eq.~\ref{Sconf2} (using $\alpha$ as the only fitting parameter; $E_0$ and $\sigma^2$ are taken 
from Fig.~\ref{Eis-fig}).}
\label{Sfigs}
\end{figure}

(i) The entropy of q-TIP4P/F water along an isochore (for constant $N$), $S(T)$, is calculated via thermodynamic
 integration; details of the calculations are given in the SM.  Fig.~\ref{Sfigs}(a) shows the $S(T)$ along selected isochores. 
While the entropy of the liquid should be positive, we find that $S(T)<0$ for all volumes considered and at $T<250-280$~K.  
This is 
 contrary to the results obtained previously for SPC/E and TIP4P/2005 water, but are consistent with the results of Habershon {\it et al.}~\cite{Habershon2011} 
using the q-TIP4P/F water model. The fact that our $S(T)<0$ is due to the bending and stretching OH bands of q-TIP4P/F water at $\omega >1400$ 
(Fig.~\ref{VDOSfig}). To show this, we note that $S=S_{conf}+ S_{vib}$ where $S_{vib}=S_{harm}(T)+S_{anharm}(T)$.  
As shown in Figs.~\ref{Sfigs}(b) and \ref{Sfigs}(c), $S_{harm}(T)<0$ while $S_{anharm}(T)>0$.  
The dominant contribution to $S$ is $S_{harm}$, which depends on the vibrational mode frequencies of the system
 (see Eqs.~\ref{ShapeF} and \ref{Svib-eqn}) and becomes increasingly more negative as the VDOS frequencies increase. 

(ii) The configurational entropy is given by $S_{conf}= S - S_{vib}$ or, equivalently, by 
$S_{conf}=S-S_{harm}(T)-S_{anharm}(T)$. The $S_{conf}(T)$ for q-TIP4P/F water is shown in 
Fig.~\ref{Sfigs}(d) for selected isochores. Interestingly, despite $S$ and $S_{harm}$ being negative,
we find that $S_{conf}>0$ (as expected, based on Eq.~\ref{SconfEqn}).  
It follows that the unphysical ($<0$) values of $S(T)$ and $S_{harm}(T)$, somehow, cancel out leading to 
a positive $S_{conf}$.

\begin{figure}[!htb]
	\centering{
		\includegraphics[width=8.0cm]{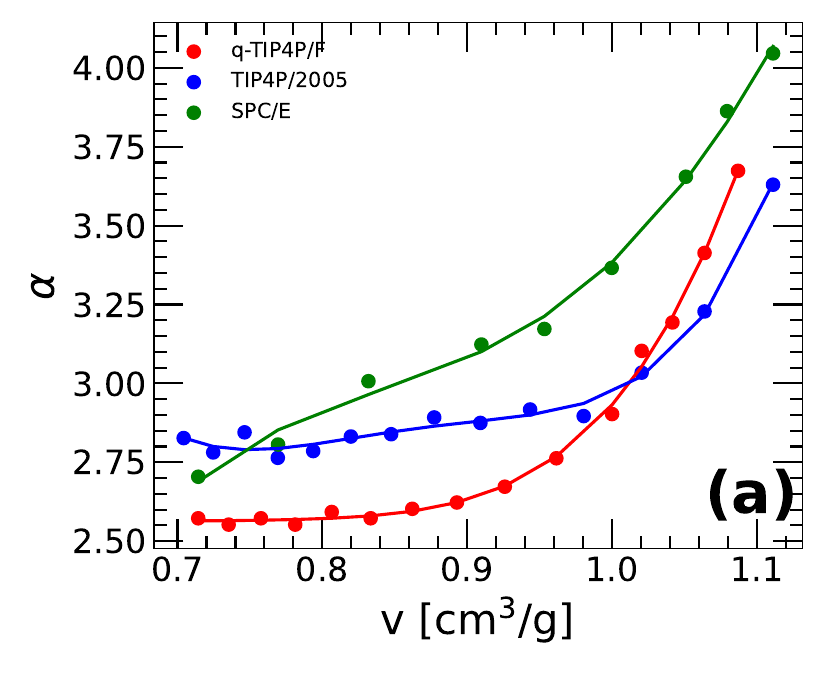}
		\includegraphics[width=8.0cm]{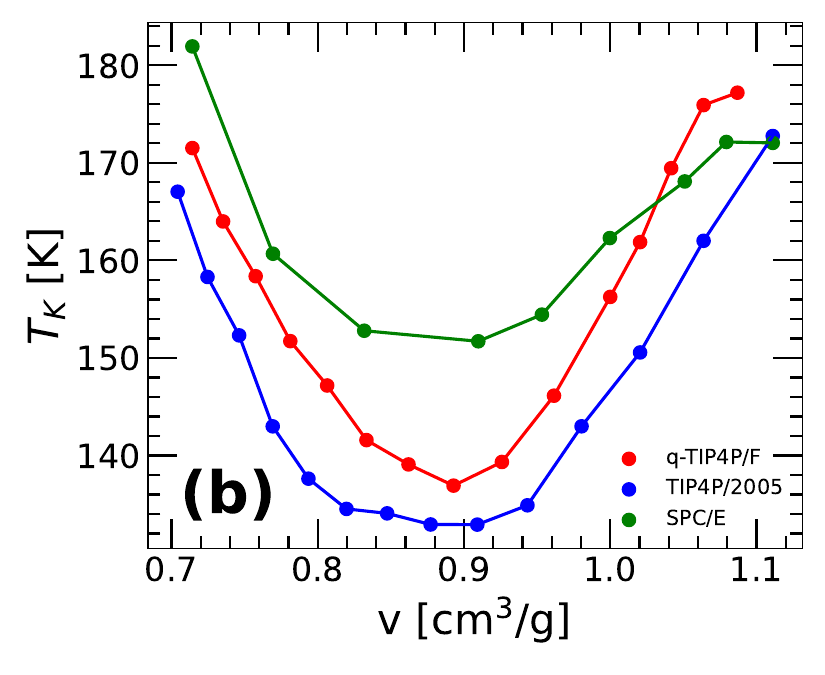}
	}
	\caption{(a) PEL parameter $\alpha(V)$ defined in Eq.~\ref{Sconf2} and obtained from the best fits shown in Fig.~\ref{Sfigs}(d) for q-TIP4P/F water. (b) Kauzmann temperature $T_K(V)$ of q-TIP4P/F water [red circles; from Eq.~29], TIP4P/2005 (blue circles), and SPC/E (green circles) water (from Refs.~\cite{handle2018potential,sciortino2003physics}). Both $\alpha(V)$ and $T_K(V)$ show a similar qualitative behavior irrespective of whether the water model is flexible (q-TIP4P/F) or rigid (TIP4P/2005 and SPC/E).}
	\label{SfigConfig}
\end{figure}

The lines shown in Fig.~\ref{Sfigs}(d) are the best fit to $S_{conf}(T)$ based on the Gaussian
approximation of the PEL, Eq.~\ref{Sconf2}. 
From these fitting curves, we extract the PEL variable $\alpha(V)$; see red circles in Fig.~\ref{SfigConfig}(a).
The values of $\alpha(V)$ obtained for q-TIP4P/F water are comparable to the corresponding values 
for  TIP4P/2005 (blue circles) and SPC/E (green circles) reported in Refs.~\cite{sciortino2003physics,handle2018potential}. 
In all three models, $\alpha(V)$ increases with increasing volume, consistent with the system having more IS 
to explore as the volume becomes larger -- the total number of IS in a Gaussian PEL is $e^{\alpha N}$. 
Surprisingly, at approximately $V<1.0$~cm$^3$/g, $\alpha(V)$ is slightly smaller for q-TIP4P/F water 
than for TIP4P/2005 and SPC/E water. This implies that including flexibility into a water model can reduce, to some degree,
 the number of IS available in the PEL, relative to the case of rigid water models. However, we also note that no PEL can
be strictly Gaussian since the Gaussian approximation of the PEL cannot hold at all temperatures. Hence, the interpretation of $e^{\alpha N}$ as the total number of IS available in the PEL should be taken with caution.

An important property of the PEL is the Kauzmann temperature $T_K(V)$.  For a given volume, $T_K(V)$ is the 
temperature at which $S_{conf}(T) = 0$~\cite{debenedetti2001supercooled}. Hence, at $T \leq T_K$, the equilibrium liquid would have only one
basin available in the PEL. It can be shown from Eq.~6 that
\begin{eqnarray}
	k_B T_K = \left( \sqrt{\frac{2\alpha N}{\sigma^2}} - b\right)^{-1}
\end{eqnarray}
The $T_K(V)$ for q-TIP4P/F water is shown in Fig.~\ref{SfigConfig}(b) (see also Fig.~\ref{Sfigs}(d)). 
At all volumes studied, the  $T_K(V)$ values for q-TIP4P/F are intermediate to the $T_K(V)$ values reported for
SPC/E and TIP4P/2005 water~\cite{sciortino2003physics,handle2018potential}.

\clearpage

\subsection{PEL-EOS of $H_2O$}
\label{phase-diagram}

\begin{figure}[!htb]
\centering{
	\includegraphics[width=7.0cm]{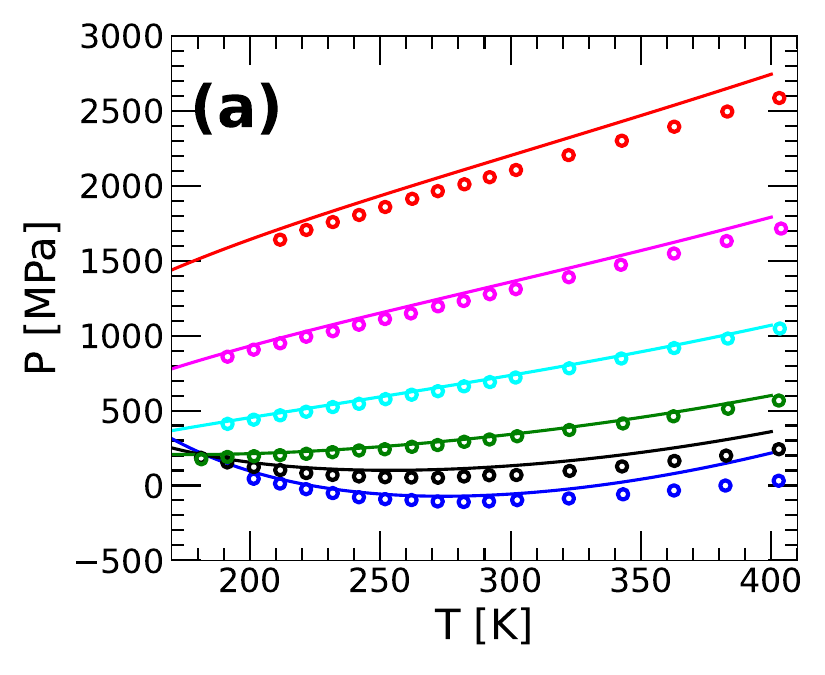}
	\includegraphics[width=7.0cm]{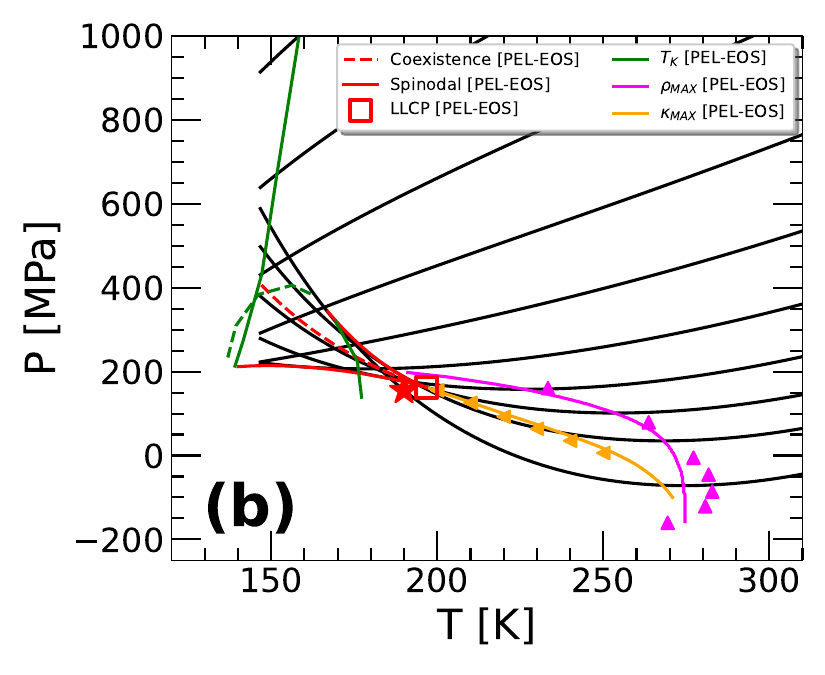}
	\includegraphics[width=7.0cm]{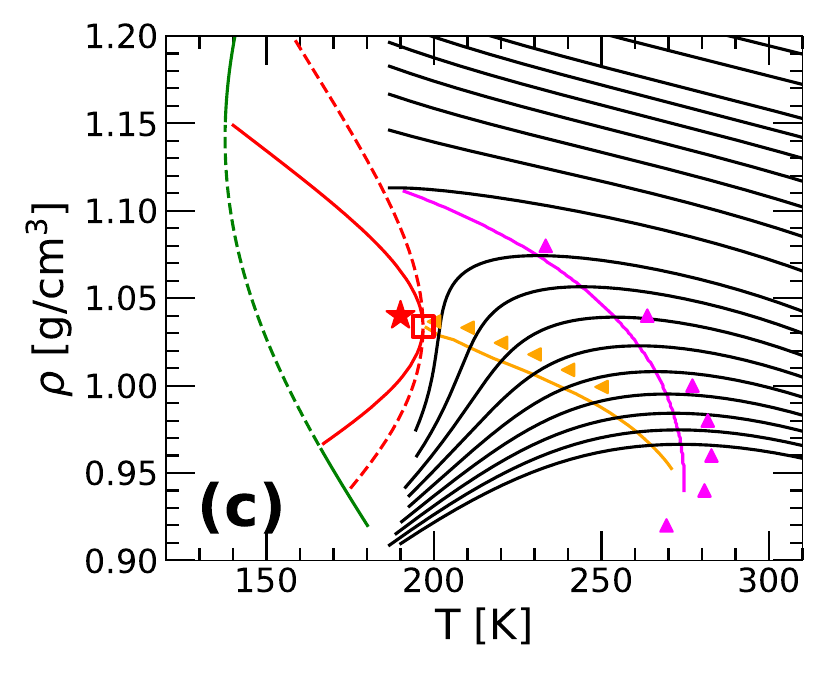}
}
\caption{(a) Isochores in the $P-T$ plane for q-TIP4P/F water obtained from MD simulations (open circles) and the PEL-EOS (solid lines) 
[Eq.~\ref{pelEos}]; $\rho = 0.96$ to $1.36$~g/cm$^3$ in steps of $0.08$~g/cm$^3$ (bottom-to-top).  
(b) $P-T$ phase diagram obtained from the PEL-EOS. Black lines are selected isobars. The red square
 is the LLCP predicted by the PEL-EOS and
the red dashed and solid lines are, respectively, the associated coexistence and spinodal lines.  Emanating from the LLCP is the associated
isothermal compressibility maxima (orange line), $\kappa_T^{max}$-line; the magenta line is the density maxima line, $\rho^{max}$-line. 
The LLCP location from MD simulations is indicated by the red star, and the corresponding $\kappa_T^{max}$-line and
$\rho^{max}$-line are indicated by orange and magenta triangles.
(c) $\rho-T$ phase diagram obtained from the PEL-EOS. Black lines are selected isobars; other lines and symbols are same as in (b).
The solid and dashed green lines in (b) and (c) are the Kauzmann temperature predicted by the PEL-EOS inside and outside the instability region.
The agreement between the PEL-EOS and MD simulations is very good, particularly at low temperatures.
}
\label{PELEOS}
\end{figure}

At this point, we calculated all the (volume-dependent) quantities of the PEL, $\{\alpha,~\sigma^2,~E_0;$ $~a,~b;~c_2,~c_3\}$, that are needed to obtain the PEL-EOS. The PEL-EOS for $P(V)$ is given by
Eq.~\ref{pelEos} and involves 
various derivatives of the PEL parameters, such as $d \alpha/dV$.  The $V$-derivatives of the parameters are 
calculated using polynomial interpolations as explained in Ref.~\cite{handle2018potential}; see also Sec.~V of the SM.  
The PEL-EOS for $P(V)$ given by Eq.~\ref{pelEos} is shown in Fig.~\ref{PELEOS}(a) (lines) together with the corresponding results 
from our MD simulations (open circles). 
For all isochores, the $P(V)$ PEL-EOS is in excellent agreement with the MD results, particular at $T\leq 280$~K. 
Deviations between the MD results and  the PEL-EOS occur at $T > 350$~K.  We note, however, that the PEL-EOS is only 
applicable at conditions where the Gaussian approximation of the PEL holds. For the case of q-TIP4P/F, the Gaussian approximations
 is applicable at $T \leq 280$~K [Fig.~\ref{Eis-fig}](a). Therefore, the good agreement between the PEL-EOS and MD simulations at
$280<T<350$~K is rather surprising. 

Figs.~\ref{PELEOS}(b) and \ref{PELEOS}(c) show, respectively, the $P-T$ and $\rho-T$ phase diagrams of q-TIP4P/F water 
obtained from the PEL-EOS. Included in these phase diagrams are the LLCP (red square), coexistence line/boundary region (red dashed-lines), 
and spinodal lines (red solid-lines), as well as the well-known anomalous maxima lines of liquid water~\cite{nilsson2015structural},
$\kappa_T^{max}$-line (orange line) and $\rho^{max}$-line (magenta line), associated to the LLCP. For comparison, we also include the LLCP, $\kappa_T^{max}$-line, and $\rho^{max}$-line obtained directly from our MD simulations.
The agreement between the PEL-EOS predictions and MD simulations is very good.  For example, deviations in the 
location of the LLCP are $\Delta T_c <10$~K, $\Delta P_c <15$~MPa, and $\Delta \rho_c <0.01$~g/cm$^3$
(the PEL-EOS predicts that $T_c = 197$~K, $P_c = 162$~MPa, and $\rho_c$ = 1.03~g/cm$^3$).

\clearpage

\subsection{Adam-Gibbs relation}
\label{adam-gibbs}

In the discussions so far, we focused on the ability of the PEL formalism 
to predict the thermodynamic properties of q-TIP4P/F water, particularly, the EOS.
In this section, we test whether there is any relationship between the dynamics of q-TIP4P/F water and its PEL.  
Indeed, it was shown in Refs.~\cite{giovambattista2003potential,handle2018adam} that the diffusion coefficient $D$ of SPC/E and TIP4P/2005 water is intimately 
related to the systems' configurational entropy via the Adam-Gibbs (AG) relationship. The AG relationship states that
\begin{equation}
        D = D_0 \exp[-A/TS_{conf}]
\label{AGeqn}
\end{equation}
where $D_0$ and $A$ are constants that depend on the particular system studied.
The AG relation is based on the vague concept of 'cooperatively rearranging regions' (CRR)~\cite{adam1965temperature}.
It follows from Eq.~\ref{AGeqn}, that $D$ decreases with decreasing $S_{conf}$, i.e., as the number of IS 
available to the system decreases.
Below we show that the AG relationship also holds for q-TIP4P/F water.

\begin{figure}[!htb]
\centering{
	\includegraphics[width=8.0cm]{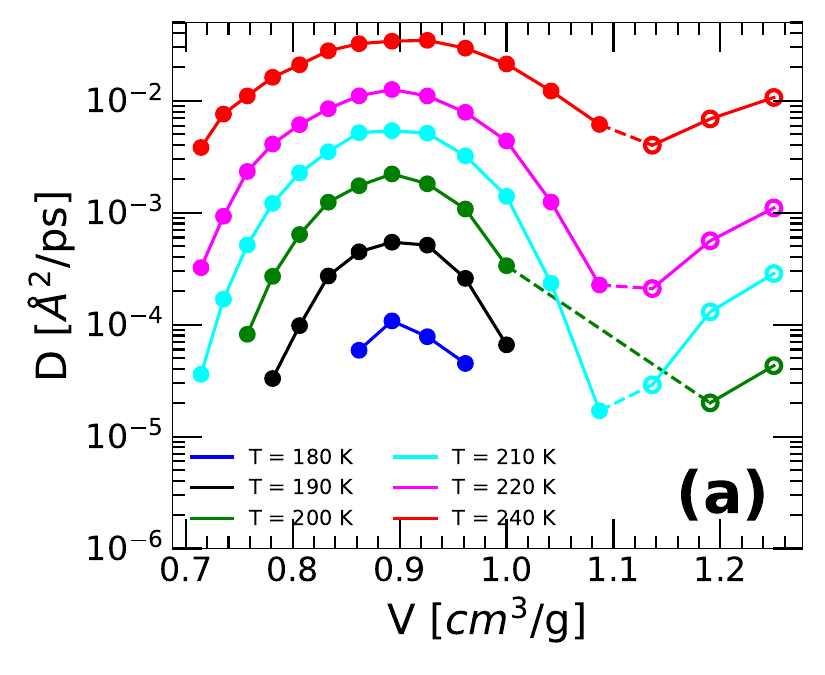}
	\includegraphics[width=8.0cm]{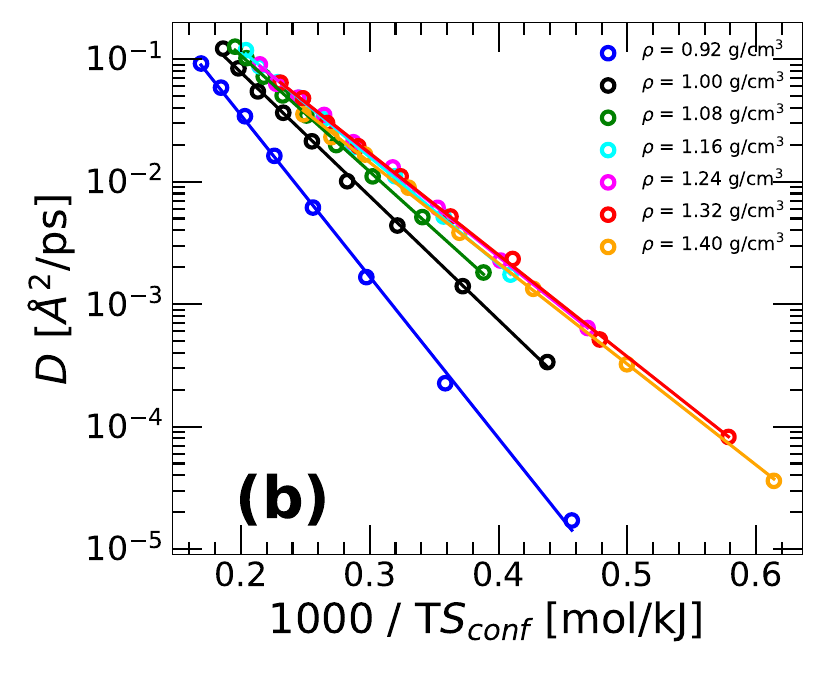}
}
\caption{(a) Diffusion coefficient $D$ of q-TIP4P/F water as a function of volume obtained from classical MD simulations at selected temperatures. 
An anomalous diffusivity maximum occurs at $V\approx 0.9$~cm$^3$/g, consistent with experiments~\cite{prielmeier1987diffusion,harris1980pressure}.  
(b) Semi-log plot of $D$ as function of  $1000/T S_{conf}$ for selected isochores. Lines in (b) correspond to the best fittings
 obtained using  the Adam-Gibbs relationship, Eq.~\ref{AGeqn}. At all volumes studied, the Adam-Gibbs relationship works remarkably well. 
}
\label{AGfig}
\end{figure}

Fig.~\ref{AGfig}(a) shows $D$ as function of volume for q-TIP4P/F water along isotherms. Consistent with computational 
and experimental results~\cite{eltareb2022evidence,harris1980pressure,prielmeier1987diffusion},
 $D(V)$ exhibits a maximum indicating that there is 
a range of volumes where the diffusion of water molecules is anomalous, i.e., $D$ decreases with increasing  $V$. 
Fig.~\ref{AGfig}(b) shows the values of $D$ [from Fig.~\ref{AGfig}(a)]
 as function of $1/T S_{conf}$ along isochores (circles) together with the corresponding predictions of the AG relation, Eq.~\ref{AGeqn}
(lines). The AG relation is fully consistent with our MD simulations.
Although our results are consistent with the previous studies of SPC/E and TIP4P/2005 water~\cite{giovambattista2003connection,handle2018adam}, it is not
evident that this should be the case.  Intuitively, one may expect that adding vibrational degrees of freedom to 
the water molecule (q-TIP4P/F model) should add roughness (and hence, more IS) to the PEL, relative to the PEL of rigid water models 
(SPC/E and TIP4P/2005 models). If so, one may expect that the $S_{conf}$ of q-TIP4P/F water should be larger than the 
$S_{conf}$ of SPC/E and TIP4P/2005 water (at a given working conditions). 
However, the values of $S_{conf}$ for all these models are of the same order of magnitude [e.g., Fig.~\ref{Sfigs}(d)].  
Indeed, the $S_{conf}$ of all three models is given by the same equation [i.e., the Gaussian approximations of the PEL, Eq.~\ref{Sconf2}], 
with very similar coefficients $\{ \alpha(V),~E_0(V),~\sigma^2(V) \}$.  It is probably the fact that all
these models have very quantitatively similar $S_{conf}(V)$, plus the fact that they all do a good job in reproducing the experimental values of $D$ for water, 
that the AG holds for the three water models considered.

\clearpage

\section{Summary and Discussion}
\label{summarySec}

In this work, we performed extensive MD simulations of water using the q-TIP4P/F model over a wide range of temperatures and volumes, and analyzed properties of the corresponding PEL by calculating the IS and the PEL curvature around the IS. Our computer simulations show that the PEL of q-TIP4P/F water is Gaussian (e.g., $E_{IS}(T)$ obeys Eq.~\ref{Eis-eqn2}) at all the densities considered in this work ($0.80 \leq \rho \leq 1.40$~g/cm$^3$) and for $T \le 280$~K (Fig.~3(a)). In addition, we find that the harmonic approximation of the PEL does not hold for q-TIP4P/F water, implying that the PEL basins contain significant anharmonicities. Nonetheless, anharmonic corrections to the PEL can be easily included by using a simple polynomial expansions (see Eq.~17 and Fig.~4).

We also calculate the configurational entropy of q-TIP4P/F water at all densities and temperatures considered. Our results indicate that $S_{conf}(T)$ is surprisingly similar to the $S_{conf}(T)$ reported for SPC/E and TIP4P/2005 water in Refs.~\cite{sciortino2003physics,handle2018potential} [see Eq.~6, Fig.~\ref{Sfigs}(d) and Fig.~\ref{SfigConfig}(b)]. This suggests that adding flexibility to the water model does not increase the number of IS available to the system (at a given N and V). We note that, even when the calculated $S_{conf}$ is physically sound and consistent with the previous results from SPC/E and TIP4P/2005 water, adding flexibility to the water model does have important implications. Indeed, consistent with Ref.~\cite{Habershon2011}, we find that the entropy of $S(T)$ of q-TIP4P/F water is negative at low temperatures. Similarly, we also find that the harmonic entropy $S_{harm}(T)$ of q-TIP4P/F water is negative. This unphysical behavior in the harmonic and total entropy of q-TIP4P/F water is due to the fact that classical statistical mechanics can not accurately describe some of the thermodynamic properties, such as the entropy, for systems with large vibrational mode frequencies. What is remarkable is that, somehow, the negative contributions to $S(T)$ and $S_{harm}(T)$ cancel out leading to a $S_{conf}(T) = S(T) - S_{harm}(T) - S_{anharm}(T)$ that is positive, and allowing for the application of the PEL formalism.

One of the main goals of this work is to show that the EOS predicted by the PEL formalism can be derived for a flexible molecular system. The obtained PEL-EOS for the case of q-TIP4P/F water depends on only seven PEL variables $\{\alpha,~\sigma^2,~E_0,~a,~b,~c_2,~c_3\}$. $E_0$ and $\sigma^2$ are related to the Gaussian approximation of the PEL and quantify the distribution of IS energies available in the system; $a$ and $b$ are related to the harmonic approximation of the PEL and quantify the curvature of the PEL about the IS; $c_2$ and $c_3$ quantify the anharmonic corrections to the PEL. The PEL-EOS calculated for q-TIP4P/F water reproduces remarkably well the corresponding EOS obtained directly from the MD simulations. In particular, the PEL-EOS predicts the LLCP to be located at ($P_c = 162$~MPa, $T_c = 197$~K, $\rho_c = 1.03$~g/cm$^3$) which is in excellent agreement with the LLCP obtained from the MD simulations, ($P_c = 150$~MPa, $T_c = 190$~K,$\rho_c = 1.04$~g/cm$^3$). Interestingly, we also find that the PEL-EOS, can be applied at $T > 280$~K, outside the range of temperatures where the Gaussian approximation holds. As shown in the SM (Table~S3-S4), the PEL formalism is robust regarding the range of $T$ and $V$ considered in the calculation of the PEL-EOS. For comparison, we note that this is not the case of the two state equation of state (TSEOS) which has been also used to predict the location of the LLCP in low temperature liquids~\cite{eltareb2022evidence,gartner2020signatures,singh2016two,holten2014two,biddle2017two,cheng2020evidence,weis2022liquid}. The location of the LLCP predicted by the TSEOS can be somewhat sensitive to the range of $T$ and $P$ considered in the calculations~\cite{gartner2022liquid,amann2023liquid}.  

In the last section of this work, we investigated the relationship between the PEL formalism and the dynamics of the system by testing the Adam-Gibbs relation for q-TIP4P/F water. The AG expression, Eq.~\ref{AGeqn}, states that the diffusion coefficient depends on the topography of the PEL, specifically $S_{conf}$. We find that the AG relation is fully consistent with our MD simulations of q-TIP4P/F water. Hence, similarly to the case for SPC/E and TIP4P/2005 water, the diffusion coefficient of q-TIP4P/F water decreases as the number of IS available to the system decreases, upon cooling.

Overall, our results for q-TIP4P/F water are consistent with previous computer simulations of water using rigid models such as SPC/E~\cite{sciortino2003physics} and TIP4P/2005~\cite{handle2018potential}. The PEL variables reported here for q-TIP4P/F are similar to the PEL variables for SPC/E and TIP4P/2005~\cite{handle2018potential,sciortino2003physics}, and show the same qualitative behavior for all densities and temperatures explored in this work. It follows that the PEL formalism can be applied to liquids/glasses composed of flexible molecules with high-frequency vibrational modes, such as q-TIP4P/F water, in the same manner as it has been applied to systems composed of rigid models.

\section{Data Availability}
The authors confirm that the data supporting the findings of this study are available within the article and its supplementary material.

\section{Acknowledgments}

This work was supported by the SCORE Program of the National Institutes of Health under award number 1SC3GM139673 and the NSF CREST 
Center for Interface Design and Engineered Assembly of Low Dimensional systems (IDEALS), NSF grant number HRD-1547380 and HRD-2112550. NG is thankful for support from the NSF, grant number CHE-2223461. AE is supported by the NSF CREST Postdoctoral Research Program, under award number 2329339. 
This work was supported, in part, by a grant of computer time from the 
City University of New York High Performance Computing Center under NSF Grants CNS-0855217, CNS-0958379 and ALI-1126113.

\section{Additional Information}
The authors declare no conflict of interest.

\section{Supporting Information}
In the SM, we provide additional information including details on (i) the crystallization process observed in the MD simulations at $\rho = 1.00$~g/cm$^3$ and low temperature, (ii) the entropy calculations, and (iii) show that the PEL-EOS reported here is robust regarding the set of $(T,V)$-thermodynamic states considered in the calculations.

\clearpage
\section{References}
\bibliography{bibliography}

\end{document}